\documentclass[11pt,document,nofootinbib,superscriptaddress,onecolumn,preprintnumbers,balancelastpage]{article}
\usepackage{jheppub} 
\usepackage{graphicx}
\usepackage{epstopdf}
\usepackage{dcolumn}
\usepackage{bm}
\usepackage{hyperref}
\usepackage{booktabs}
\usepackage[dvipsnames]{xcolor}
\usepackage{amsmath}
\usepackage{amssymb}
\usepackage{cancel}
\usepackage{xpatch}
\usepackage{fontawesome} 
\usepackage{arydshln}
\usepackage{xspace}
\usepackage{mathtools}
\usepackage{bbold}
\usepackage{booktabs, multirow} 
\usepackage{soul}
\usepackage{tikz}
\usepackage{graphicx} 
\usepackage{physics}
\usepackage{comment}
\usepackage[shortlabels]{enumitem}

\renewcommand{\dd}{{\rm d}}
\renewcommand{\sec}[1]{section~\ref{#1}}

\definecolor{deepgreen}{rgb}{0.2,0.8,0.2}

\definecolor{deepblue}{rgb}{0.2,0.4,0.8}

\definecolor{deepred}{rgb}{0.8,0.2,0.2}

\usepackage{todonotes}

\newcommand{\githubmaster}{\href{https://github.com/mkongsore/DyonDescender}{\faGithub}\xspace}

\allowdisplaybreaks

\title{Dyon Loops and Abelian Instantons}
\author[a]{Isabel Garcia Garcia,}
\author[b]{Marius Kongsore,}
\author[b,c]{and Ken Van Tilburg}

\affiliation[a]{Department of Physics, University of Washington, Seattle, WA 98195, USA}
\affiliation[b]{Center for Cosmology and Particle Physics, Department of Physics,
New York University, New York, NY 10003, USA}
\affiliation[c]{Center for Computational Astrophysics, Flatiron Institute, New York, NY 10010, USA}

\emailAdd{isabelgg@uw.edu}
\emailAdd{mkongsore@nyu.edu}
\emailAdd{kenvt@nyu.edu}

\abstract{We construct Abelian gauge field configurations that carry non-zero instanton number. Each such ``Abelian instanton" is generated by a closed magnetic worldline in four-dimensional Euclidean space, provided the Abelian gauge field has non-trivial winding along the closed worldline. The resulting field configuration corresponds to a Euclidean dyon loop featuring non-zero instanton number. We embed these dyon loops in a UV-complete theory using the Georgi-Glashow model and show that the full instanton charge is borne entirely by the unbroken $U(1)$ sector. In this same model, using a numerical relaxation procedure, we show that Euclidean dyon loops are a continuous deformation of small BPST instantons.}

\begin{document}
\maketitle
\flushbottom

\section{Introduction}\label{sec:intro}

Monopoles and instantons are non-perturbative, topological field configurations that have been crucial for understanding the structure and dynamics of gauge theories. Magnetic monopoles,  proposed by Dirac~\cite{Dirac:1931kp} and explicitly constructed by 't~Hooft~\cite{tHooft:1974kcl} and Polyakov~\cite{Polyakov:1974ek}, generically arise in extensions of the Standard Model: from grand unified theories, to extra-dimensional constructions, to string theory.
Their presence is so generic that it led Polchinski to formulate his completeness hypothesis~\cite{Polchinski:2003bq}, and cosmic inflation was introduced (in part) of get rid of all the monopoles that would be produced as a result of spontaneous symmetry breaking in the early universe~\cite{Kibble:1976sj}. In the presence of a vacuum angle $\theta$, a magnetic monopole turns into a dyon with a $\theta$-dependent electric charge~\cite{Witten:1979ey} --- the first demonstration of how the $\theta$ angle can affect physical observables related to magnetic monopoles. Instantons, finite-action solutions to the Euclidean Yang-Mills equations, mediate tunneling processes and contribute non-perturbatively to the breaking of axial symmetries and the vacuum susceptibility of gauge theories.

In this work, we establish a novel connection between these two classes of topological objects, by identifying and describing Abelian gauge field configurations that carry non-zero instanton number. These ``Abelian instantons" correspond to monopole loop configurations in four-dimensional Euclidean space that feature non-trivial winding of the gauge field $A$ along the loop, and can be described as Euclidean dyon loops featuring localized magnetic charge but slightly ``smeared" electric charge density. Due to the singular nature of Dirac monopoles, a $U(1)$ gauge theory with a monopole loop is not defined in all of $\mathbb{R}^4$ but rather $\mathbb{R}^4 \setminus S^1$. The latter is a topologically non-trivial space featuring an internal boundary $S^1 \times S^2$ that surrounds the monopole loop. Simultaneous winding of $A$ along the $S^1$ and wrapping of $F=\dd A$ around the $S^2$ leads to a non-zero instanton number. We study the UV-complete equivalent of these Abelian dyon loops in the Georgi-Glashow model, i.e.~$SU(2)$ gauge theory spontaneously broken to $U(1)$ via an adjoint Higgs field. Implementing a numerical relaxation algorithm, we show that a Euclidean loop of the minimally charged Julia-Zee dyon is a continuous deformation of the familiar (small) non-Abelian instanton present in the Georgi-Glashow model~\cite{Agasian:2014uua}. Furthermore, we show that the instanton number associated with these non-Abelian dyon loops is carried entirely by the unbroken Abelian component. Specifically, in singular gauge, where the Abelian and $W$ gauge field degrees of freedom are fully separated, the instanton number can be computed from just the unbroken Abelian gauge field component, allowing us to relate these non-Abelian instantons to their infrared remnants.

The Abelian instantons we present in this work are the explicit field configurations that capture the physical effect first described in ref.~\cite{Fan:2021ntg}, which used a dyon worldline effective action to argue that radiative corrections from monopole loops in a $U(1)$ gauge theory induce a $\theta$-dependent contribution to the vacuum energy density. Specifically, the $\theta$-term converts monopoles into dyons with masses dependent on $\theta$~\cite{Witten:1979ey}, in turn generating radiative corrections to the vacuum energy that are sensitive to $\theta$. If $\theta$ is a dynamical axion field, ref.~\cite{Fan:2021ntg} argued this effect would amount to a contribution to the axion potential, with possible implications for the QCD axion as a solution to the strong CP problem. 

The connection between monopoles and instantons in non-Abelian gauge theories has been the subject of longstanding interest. 
A free Julia-Zee dyon can carry instanton number in a spacetime with a compact time direction, e.g.~$\mathbb{R}^3 \times S^1$~\cite{Marciano:1976as,Christ:1979iw,Kim:1980rv}. Similarly, a static dyon worldline in $\mathbb{R}^4$ is equivalent to an infinite sequence of instantons~\cite{Rossi:1978qe,Rossi:1982fq}. Taubes proved the existence of unstable monopole--antimonopole bound states~\cite{Taubes:1984scl}, using Morse theory to connect the existence of saddle points in the sector of vanishing magnetic charge to the existence of non-contractible loops in configuration space corresponding to isospin-twisted magnetic monopoles. Refs.~\cite{Jahn:1999wx,Bruckmann:2000zs,Bruckmann:2002jm} elaborated on Taubes' construction by noting connections between the topology of instantons and that of dyons, while ref.~\cite{Saurabh:2017ryg} numerically studied the interactions of monopole--antimonopole pairs under this ``Taubes twist'' and found the explicit field configuration of the metastable bound state. Motivated by understanding Yang-Mills confinement, circular monopole current loops (defined via maximal Abelian projection) have also been investigated in instanton backgrounds in the context of unbroken $SU(2)$ gauge theory~\cite{Chernodub:1995tt,Hart:1995wk,Brower_1997,Bornyakov:1997at}. Recently, infinite monopole worldlines carrying Abelian instanton number have been found to lead to physical effects on fermion fields, related to the Callan-Rubakov effect~\cite{Csaki:2024ajo}.
A concurrent publication~\cite{Chen:2025buv} reveals the existence of a more general class of $\theta$-terms in a wide class of field theories, including Maxwell theory and beyond.

Dyon loops have also appeared in the context of generalized symmetries in quantum field theory. Most notably, whenever a zero-form chiral symmetry or axion shift symmetry is broken by an ABJ anomaly~\cite{Adler:1969gk,Bell:1969ts}, the (broken) symmetry current is sourced by an Abelian $F \wedge F$ term~\cite{Choi:2022fgx,PhysRev.177.2426,Bell:348417}.  However, no smooth Abelian field configuration charged under $F \wedge F$ exists in a topologically trivial spacetime --- standard lore says that ``there are no Abelian instantons". Nevertheless, a consequence of ref.~\cite{Fan:2021ntg} is that monopoles \emph{must} be able to play the role of exactly such instantons. One particularly elegant way of understanding this is in terms of non-invertible symmetries: the anomalous axion shift symmetry (or alternatively, chiral symmetry) is not completely broken, but has a non-invertible symmetry remnant that nonetheless acts invertibly on the axion (chiral fermion) operator~\cite{Choi:2022fgx,shao2024whatsundonetasilectures}. The non-invertibility appears when the symmetry acts on 't Hooft lines --- suggesting that dynamical monopoles, which lead to the enability of 't Hooft lines, play a role in violating the selection rules associated with the symmetry non-perturbatively~\cite{Cordova:2022ieu,Choi:2022fgx,shao2024whatsundonetasilectures}.

While these previous studies establish important links between monopoles/dyons and instantons, we show in this work that:
\begin{enumerate}[(i)]
    \item Pure $U(1)$ gauge theory can support finite instanton number on $\mathbb{R}^4$ in the form of (singular) Abelian gauge fields corresponding to closed loops of dyon worldlines; 
    \item The instanton number associated with a dyon worldline (open or closed) is entirely carried by the unbroken Abelian component, and can thus be described by the IR degrees of freedom in the effective $U(1)$ theory; 
    \item Dyon loops in the Georgi-Glashow model are continuous deformations of the familiar ``small" or ``constrained" instanton solutions present in the Higgsed phase~\cite{Affleck:1980mp}, as was originally conjectured by ref.~\cite{Fan:2021ntg}.
\end{enumerate}
We suspect that these novel findings may open the door to new ways of studying instanton effects, especially when the exact details of the dyon loop UV completion are unknown.

The remainder of this paper is organized as follows. In~\sec{sec:abelian-loop}, we identify Abelian gauge field configurations that carry instanton number, and show that they can be described as dyon loops in $\mathbb{R}^4$.
Section~\ref{sec:free-dyon} begins with a review of the Julia-Zee dyon solution of the Georgi-Glashow model and its topological aspects in $\mathbb{R}^3 \times S^1$. We then show that the instanton number of the free Julia-Zee dyon, as originally identified in refs.~\cite{Marciano:1976as, Rossi:1978qe, Christ:1979iw, Kim:1980rv,Rossi:1982fq} in its non-Abelian form, is entirely carried by the unbroken Abelian component.
In \sec{sec:non-abelian-loop}, we study the Georgi-Glashow UV-complete equivalents of the Abelian field configurations of \sec{sec:abelian-loop}, and establish their relation to the constrained BPST instanton that survives Higgsing. We conclude in \sec{sec:conclusions} and comment on future directions and applications. Appendix~\ref{sec:app_action} presents the steps in the calculations of the $U(1)$ instanton action summarized in~\sec{sec:u1action}. Our numerical relaxation procedure is detailed in appendix~\ref{app:numerical_relaxation}; the associated code is available on GitHub~\githubmaster. In appendix~\ref{app:bpst_hopf_map}, we explicitly match the topology of the BPST instanton at infinity to that of a dyon loop --- a result used in~\sec{sec:non-abelian-loop}. Finally, Appendix~\ref{app:no-other-action-minimum} contains an analytic proof that the zero-size BPST instanton is the unique exact solution to the classical Euclidean Georgi-Glashow equations of motion.

\paragraph{Notation and conventions  ---} We use the term ``instanton'' to refer to field configurations that carry non-zero instanton number (or equivalently, second Chern number or Pontryagin index). This includes gauge field configurations that are not exact solutions to the Euclidean equations of motion but that nevertheless feature non-zero $\int F \wedge F$. As in the case of so-called ``constrained instantons''~\cite{Affleck:1980mp}, we solve the equations of motion subject to a constraint, in this case the presence of a monopole loop at finite radius.

We operate exclusively within Euclidean space throughout this work. In \sec{sec:abelian-loop}~and~\ref{sec:non-abelian-loop}, we will find it convenient to chart $\mathbb{R}^4$ with double polar coordinates $\{ u, \varphi, v, \tau \}$, related to the familiar Cartesian coordinates $\lbrace x_\mu \rbrace$ as follows
\begin{equation} \label{eq:coords_dp}
    u e^{i \varphi} \equiv x_1 + i x_2 \qquad \text{and} \qquad v e^{i \tau} \equiv x_3 + i x_4 \ ,
\end{equation}
with $u, v \in [0, \infty)$ and $\varphi, \tau \in [0,2\pi)$.
Occasionally, we also use toroidal coordinates $\{ \lambda, \varphi, \vartheta, \tau \}$, which are a natural coordinate system in a space containing circular line defects. The azimuthal angles $\varphi$ and $\tau$ are as defined in eq.~\eqref{eq:coords_dp}, whereas $\lambda \in [0, \infty)$ and $\vartheta \in [0, \pi]$ are radial and angular coordinates, defined implicitly in terms of $u$ and $v$ as follows
\begin{equation} \label{eq:coords_tor}
    u = \frac{R \sin \vartheta}{\cosh \lambda - \cos \vartheta} \qquad \text{and} \qquad v = \frac{R \sinh \lambda}{\cosh \lambda - \cos \vartheta} \qquad \text{for} \qquad R \geq 0 \ .
\end{equation}
In these coordinates, $\lambda \rightarrow \infty$ corresponds to $u=0$ and $v=R$, i.e.~a loop of radius $R$ in the $x_3 x_4$ plane. The loop itself is parameterized by $\tau \in [0, 2\pi)$, with the angles $\varphi$ and $\vartheta$ left undefined.

We work with non-canonically normalized gauge fields, so that a factor of $1/e^2$ appears in front of the kinetic term of the action
\begin{subequations}
\label{eq:action_abelian}
\begin{align}
S_E &=  \int \left\lbrace \frac{1}{2e^2} F \wedge \star F + i \frac{\theta}{8\pi^2} F \wedge F \right\rbrace  \\
&= \int \dd^4 x \, \left\lbrace \frac{1}{4e^2} F^{\mu \nu} F_{\mu \nu} + i \frac{\theta}{32\pi^2} \epsilon^{\mu \nu \rho \sigma} F_{\mu \nu} F_{\rho \sigma} \right\rbrace \ ,
\end{align}
\end{subequations}
with $e$ the corresponding gauge coupling, and $\theta$ the vacuum angle. We adopt a similar normalization for the Georgi-Glashow model.

The electric and magnetic charges carried by excitations are determined by integrals of the associated conserved 1-form currents $j_e$ and $j_m$, respectively. In the normalization adopted here, $J_e \equiv \star j_e$ is a conserved 3-form, $\mathrm{d} J_e=0$, which couples to the gauge potential $A$ through the interaction term $S_\mathrm{int} \supset \int A \wedge J_e$. Likewise, magnetic charges are sourced by a conserved 3-form $J_m \equiv \star j_m$, which modifies the Bianchi identity for $F$. The magnetic current couples (in a dual picture) to a magnetic potential $A_m$, with the coupling $S_\mathrm{int} \supset \int A_m \wedge J_m$.

In the presence of both electric and magnetic charges, the Euclidean Maxwell equations in this normalization read
\begin{subequations}
\label{eq:maxwell}
\begin{align}
  \frac{1}{e^2}\,\mathrm{d}\star F &= J_e\,;\label{eq:maxwell_inhom} \\
  \mathrm{d} F &= J_m\,. \label{eq:maxwell_bianchi}
\end{align}
\end{subequations}
These enforce conservation of the respective currents, $\mathrm{d} J_e=\mathrm{d} J_m=0$.
The corresponding electric and magnetic charges $q_e$ and $q_m$, enclosed by a 3-dimensional volume $\mathcal{V}_3$ bounded by a surface $\Sigma_2 = \partial \mathcal{V}_3$, are
\begin{subequations}
\begin{alignat}{7}
q_e &= \int_{\mathcal{V}_3} J_e 
&&= \frac{1}{e^2} \int_{\Sigma_2}* F \;\in\;\mathbb{Z} \,; \\
q_m &= \int_{\mathcal{V}_3} J_m
&&= \int_{\Sigma_2} F 
\;\in\; 2\pi \, \mathbb{Z} \,.
\end{alignat}
\end{subequations}
The presence of the factor $1/e^2$ in eqs.~\eqref{eq:action_abelian}~\&~\eqref{eq:maxwell_inhom} implies that the \emph{physical electric charge} is given by $Q_e = e\,q_e$, whereas the \emph{physical magnetic charge} is $Q_m = q_m/e$. This ensures that the Dirac quantization condition takes the standard form:
\begin{equation}
Q_e Q_m = q_e q_m \;\in\;2\pi\,\mathbb{Z} \,.
\end{equation}

\section{Dyon Loops as Abelian Instantons}\label{sec:abelian-loop}

The results of ref.~\cite{Fan:2021ntg} suggest that in the presence of magnetic monopoles, there exist Abelian gauge field configurations that carry non-zero instanton number. At first, this statement may seem puzzling. In $\mathbb{R}^4$, instantons are in one-to-one correspondence with maps from the infinite spatial three-sphere to the vacuum manifold $\mathcal{M}$, and are therefore classified by the third homotopy group of $\mathcal{M}$.\footnote{It is often useful to think of non-Abelian instantons as existing on $S^4$, the one-point compactification of $\mathbb{R}^4$, although we do not utilize this notion in this work.} But since $U(1)\cong S^1$ and $\pi_3 (S^1) = \mathbb{1}$, no Abelian instantons exist on a topologically trivial spacetime. This leaves open the possibility of Abelian instantons on spacetimes with non-trivial topology, the canonical example being a configuration defined on a four-torus so that electric and magnetic flux independently wrap two of the two-tori~\cite{Reece:2023czb}.

The Abelian instantons that are the focus of this work belong to a separate category. Asymptotically, the spacetime has the geometry of $\mathbb{R}^4$, but the $U(1)$ gauge theory is not well defined on \emph{all} of $\mathbb{R}^4$ due to the presence of magnetic monopoles. Specifically, closed magnetic worldlines modify the spacetime on which the theory is defined, turning it into a manifold with non-trivial topology and a three-dimensional boundary surrounding the monopole worldline. In the low-energy effective theory, the Abelian gauge field can wind non-trivially around this boundary, giving rise to a quantized Abelian instanton number.

The rest of this section is organized as follows. We begin in \sec{sec:u1monopole} by describing the field configuration corresponding to a Dirac monopole traveling on a circular worldline, and identify the necessary requirements for this field configuration to carry non-zero instanton number. In \sec{sec:u1dyon}, we show that this Abelian instanton describes a monopole loop that is accompanied by a smooth distribution of electric charge. The resulting object is a closed worldline of a dyon with point-like magnetic charge and localized (but slightly ``smeared'') electric charge. We estimate the corresponding instanton action in \sec{sec:u1action}.

\subsection{Abelian Instantons on Closed Magnetic Worldlines}
\label{sec:u1monopole}

\begin{figure}
    \centering
    \includegraphics[scale=1.00]{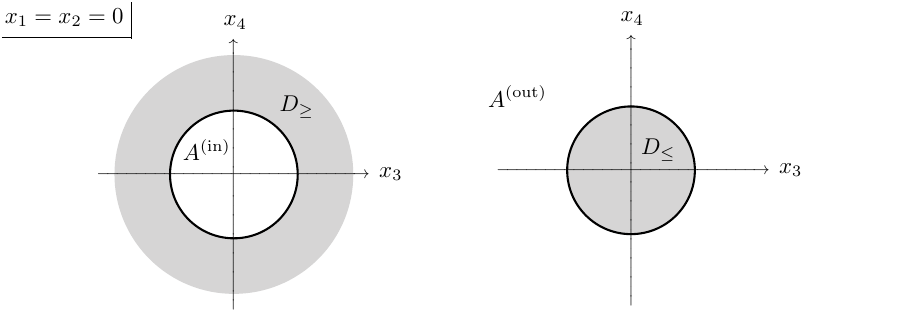}
    \caption{The one-form gauge fields $A^\text{(in)}$ and $A^\text{(out)}$, given by Eqs.\,\eqref{eq:Ain} and \eqref{eq:Aout}, are singular on the two-dimensional Dirac sheets $D_\geq = \{ u=0, v \geq R \}$ and $D_\leq = \{ u=0, v \leq R \}$, respectively. Their respective domains are therefore $\mathcal{M}^\text{(in)} = \mathbb{R}^4 \, \backslash \, D_\geq$ and $\mathcal{M}^\text{(out)} = \mathbb{R}^4 \, \backslash \, D_\leq$. In the intersecting region $\mathcal{M}^\text{(in)} \cap \mathcal{M}^\text{(out)}$, $A^\text{(in)}$ and $A^\text{(out)}$ are equivalent up to the singular gauge transformation given by eq.~\eqref{eq:gauge_phi}.} 
    \label{fig:sheets}
\end{figure}

We now explicitly construct a Dirac magnetic monopole loop field configuration and discuss the conditions under which it can carry instanton number. As a starting point, we consider a magnetic one-form current oriented counterclockwise along a circular loop in Euclidean space, corresponding to a closed monopole worldline. This current is given by
\begin{subequations} \label{eq:jmonopoleloop}
\begin{align} 
    j_m & = q_m \delta(x_1) \delta(x_2) \delta(v-R) \left(-\sin \tau \, \dd x_3 + \cos \tau \, \dd x_4 \right) \label{eq:jmonopoleloopv1}\\
        & = 2 m \frac{v}{u} \delta(u) \delta(v-R) \, \dd\tau \ , \label{eq:jmonopoleloopv2}
\end{align}
\end{subequations}
with $q_m = 2\pi m \in 2\pi \, \mathbb{Z}$ the monopole charge and $R$ the loop radius. Without loss of generality, we have chosen the monopole loop to lie in the $x_3x_4$ plane. As the angular coordinate $\tau$ varies from $0$ to $2 \pi$, the point-like monopole travels along the full circle.
In turn, the electromagnetic two-form field strength that satisfies Maxwell's equations in the presence of this magnetic current is given by
\begin{align}
    F &= - 2 m \frac{R^2 u}{d(u,v)^{3/2}} \Big[ (R^2 + u^2 - v^2) \, \dd u \wedge \dd \varphi + 2 u v \, \dd v \wedge \dd \varphi \Big] \ ; \label{eq:Fmonopoleloop} \\
    d(u,v) &\equiv \left[ u^2 + (v-R)^2 \right] \left[ u^2 + (v+R)^2 \right] \ . \label{eq:duv}
\end{align}
The field strength of eq.~\eqref{eq:Fmonopoleloop} is well defined globally and only features a (physical) singularity at $\{u=0,\,v=R\}$. However, just like for the static Dirac monopole, the corresponding one-form gauge field can only be defined \emph{locally}~\cite{Dirac:1931kp,Wu:1975es}.
For instance, consider the following vector potential
\begin{equation} \label{eq:Ain}
    A^{\text{(in)}} = - \frac{m}{2} \left[ \frac{u^2 + v^2 - R^2}{d(u,v)^{1/2}} + 1 \right] \dd \varphi \ ,
\end{equation}
with $d(u,v)$ from eq.~\eqref{eq:duv}. It is easy to check that this vector potential yields the correct field strength $F = \dd A^{\text{(in)}}$. But notice that
\begin{align}
    \lim_{u \rightarrow 0} A^{\text{(in)}} = - \frac{m}{2} \left[ \frac{v - R}{|v-R|} + 1 \right] \dd \varphi
    = \left. \begin{cases} - m & \text{for} \quad v > R \\ 0 & \text{for} \quad v < R \end{cases} \right\} \dd \varphi \ ,
\end{align}
indicating the presence of a gauge field singularity at $\{u=0,\,v>R\}$ since $\dd\varphi$ blows up there. So provided $m\neq 0$, the gauge field $A^{\text{(in)}}$ has an unphysical singularity at $u=0$ ``outside'' the monopole loop, as well as a physical singularity on the monopole loop itself.
$A^{\text{(in)}}$ is therefore only a smooth, well-defined gauge field away from the two-dimensional surface $D_\geq = \{ u=0, v \geq R \}$. This surface, $D_\geq$, is a so-called ``Dirac sheet'': the two-dimensional worldsheet traced by the monopole's Dirac string as the monopole travels along its worldline.
As with the familiar Dirac string, the location of this Dirac sheet is not physical. For example, consider instead the following one-form gauge field
\begin{equation} \label{eq:Aout}
    A^{\text{(out)}} = - \frac{m}{2} \left[ \frac{u^2 + v^2 - R^2}{d(u,v)^{1/2}} - 1 \right] \dd \varphi \ .
\end{equation}
Like before, $F= \dd A^{\text{(out)}}$, but we now have
\begin{equation}
    \lim_{u \rightarrow 0} A^{\text{(out)}} = - \frac{m}{2} \left[ \frac{v - R}{|v-R|} - 1 \right] \dd \varphi
    = \left. \begin{cases} 0 & \text{for} \quad v > R \\ m & \text{for} \quad v < R \end{cases} \right\} \dd \varphi \ ,
\end{equation}
so the Dirac sheet now coincides with a closed disk~$D_\leq = \{ u=0, v \leq R\}$ bounded by the monopole loop. We illustrate each of the two Dirac sheet configurations in figure~\ref{fig:sheets}. $A^\text{(in)}$ and $A^\text{(out)}$ are regular everywhere away from their corresponding Dirac sheets, which we take to be excluded from their respective domains. In each corresponding domain, the exterior derivative yields the two-form field strength of eq.~\eqref{eq:Fmonopoleloop}. On the intersection of the domains, the two gauge fields are related by the following gauge transformation
\begin{align}
    A^\text{(out)} = A^\text{(in)} + i g \dd g^{-1},  \qquad g = e^{im\varphi} \ , \label{eq:gauge_phi}
\end{align}
which itself is singular in the $u=0$ plane.

Given its importance, it is worth restating the content of the previous paragraph from a slightly different perspective. In the toroidal coordinate system introduced in eq.~\eqref{eq:coords_tor}, the two-form field strength of eq.~\eqref{eq:Fmonopoleloop} reads
\begin{equation} \label{eq:Fmonopoleloop_tor}
    F = \frac{m}{2} \sin \vartheta \, \dd \vartheta \wedge \dd \varphi \ ,
\end{equation}
and the corresponding one-form gauge potentials of eqs.\,\eqref{eq:Ain} and \eqref{eq:Aout} are given by
\begin{equation} \label{eq:Amonopoleloop_tor}
    A^{\text{(in)}} = - \frac{m}{2} (\cos \vartheta + 1) \dd \varphi \qquad \text{and} \qquad A^{\text{(out)}} = - \frac{m}{2} (\cos \vartheta - 1) \dd \varphi \ .
\end{equation}
In these coordinates, the field strength and gauge field components take the exact same form as those of the static Dirac monopole.
Per eq.~\eqref{eq:coords_tor}, $\vartheta = 0$ corresponds to $\{u=0,\, v>R\}$, and $\vartheta = \pi$ corresponds to $\{u=0,\, v<R\}$.
Since $A^{\text{(in)}} \Big|_{\vartheta = 0} \neq 0$ and $A^{\text{(out)}} \Big|_{\vartheta = \pi} \neq 0$, their corresponding domains must exclude the two-dimensional surfaces outside and inside of the monopole loop, respectively. The monopole loop itself corresponds to the limit $\lambda \rightarrow \infty$ with $\vartheta$ left undefined. $A^{\text{(in)}}$ and $A^{\text{(out)}}$ are therefore both ill-defined on the monopole worldline, so the circle must be excised from their respective domains, leaving $\mathbb{R}^4 \setminus S^1$.

In complete analogy with the Wu-Yang construction of the static Dirac monopole \cite{Wu:1975es}, a non-singular description of the magnetic loop requires that we give up on the standard parametrization of the space surrounding the loop by a single set of coordinates, and instead divide the space into two overlapping submanifolds. We can define the one-form gauge field using two charts, as follows:
\begin{equation} \label{eq:Aloop}
    A = \begin{cases}
    A^\text{(in)} & \forall \ x \in \mathcal{M}^\text{(in)} = \mathbb{R}^4 \, \backslash \, D_\geq \\
    A^\text{(out)} & \forall \ x \in \mathcal{M}^\text{(out)} = \mathbb{R}^4 \, \backslash \, D_\leq \end{cases} \ .
\end{equation}
This defines a vector potential that is non-singular everywhere in the union of the two domains, i.e.~$\mathcal{M}^\text{(in)} \cup \mathcal{M}^\text{(out)} = \mathbb{R}^4 \, \backslash \, S^1$. The  base space of a $U(1)$ gauge theory with a monopole loop is therefore not $\mathbb{R}^4$, but $\mathbb{R}^4$ with the loop removed. The non-trivial topology of this underlying manifold is the crucial ingredient that allows the theory to support instantons.

\begin{figure}
    \centering
    \includegraphics[width=0.9\textwidth, trim=10 10 10 -4pt, clip]{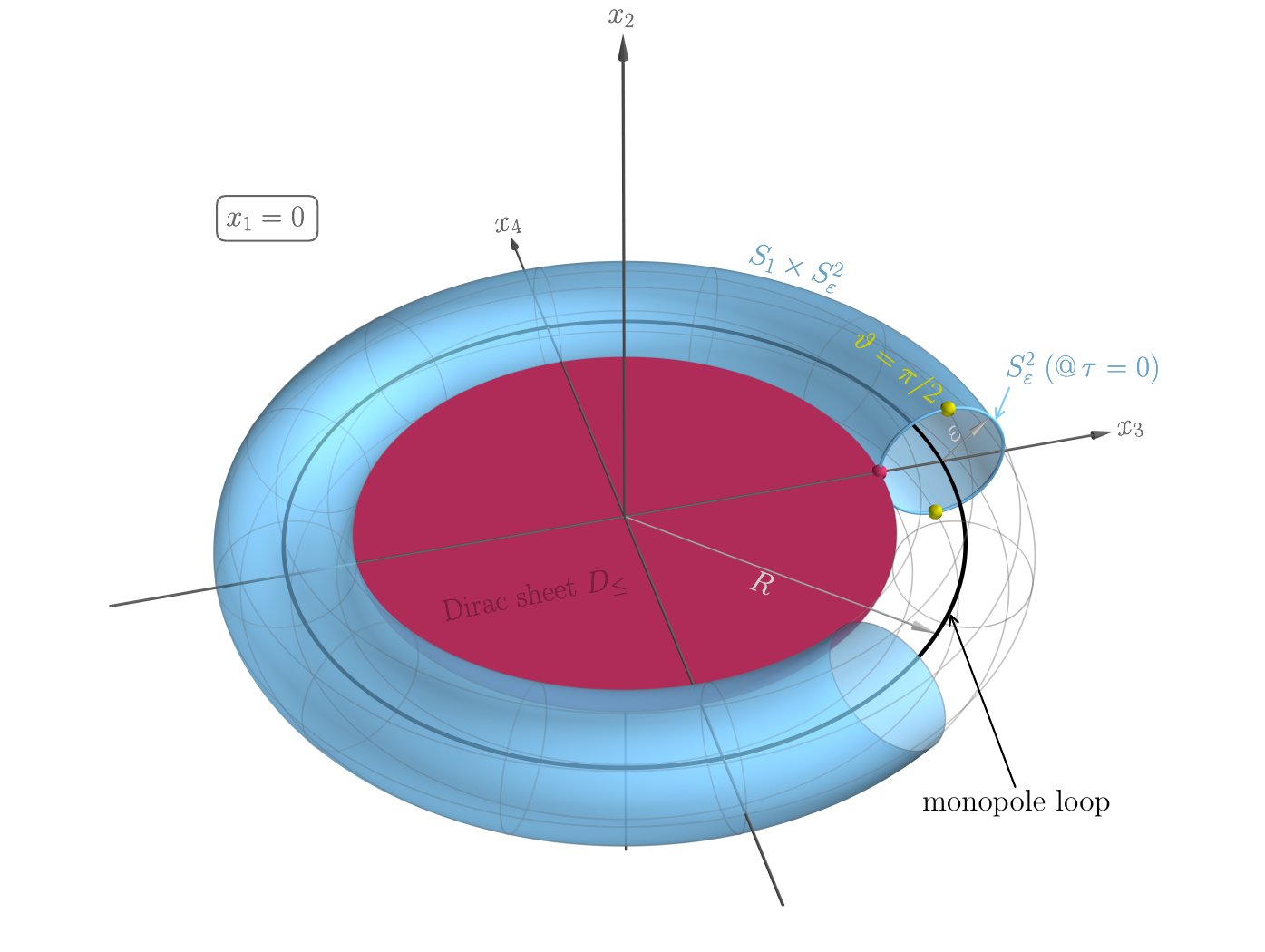}
    \caption{Depiction of the closed monopole worldline (black loop) at $u = \sqrt{x_1^2 + x_2^2} = 0$ and radius $v=\sqrt{x_3^2 + x_4^2}=R$, parametrized by $\tau \in [0,2\pi)$ (see eq.~\eqref{eq:coords_dp}). The blue circle represents the 2-sphere $S^2_\varepsilon$ of radius $\varepsilon$ around the monopole at $\tau = 0$; only the slice at $x_1 = 0$ is shown. The Abelian gauge field of eq.~\eqref{eq:Aout} is singular at the south pole ($\vartheta = \pi$, red point) of this $S^2_\epsilon$, and generally everywhere along the Dirac sheet $D_\leq$ (red disk), the 2-surface defined by $\lbrace u = 0, v\leq R \rbrace$. The $S^1_\varepsilon$ equator of the $S^2_\varepsilon$ at $\tau=0$ is represented by the two yellow points at $\vartheta = \pi/2$. The monopole loop $S^1$ is completely enclosed by the boundary 3-surface $S^1 \times S^2_\varepsilon$ (blue torus) defined by $u^2 + (v-R)^2 = \varepsilon^2$. } 
    \label{fig:summary}
\end{figure}

To go further, it is convenient to consider the monopole to have a finite radius $\varepsilon$ (at fixed $\tau$). In this description, the monopole loop carves out the four-volume $S^1 \times B^3_\varepsilon$ from our Euclidean space, making our remaining spatial manifold $\mathbb{R}^4 \, \backslash \, \{ S^1 \times B^3_\varepsilon \}$ instead of $\mathbb{R}^4 \, \backslash \, S^1$. The latter is recovered in the limit $\varepsilon \rightarrow 0$. The manifold $\mathbb{R}^4 \, \backslash \, \{ S^1 \times B^3_\varepsilon \}$ has two boundaries: an external boundary $S^3_\infty$ (the familiar 3-sphere ``at infinity"), and an internal boundary $S^1 \times S^2_\varepsilon$. We plot a slice (at $x_1=0$) of this internal boundary 3-surface in figure~\ref{fig:summary}.
The internal boundary is naturally parametrized by the three angular variables of our toroidal coordinate system $\{ \lambda, \varphi, \vartheta, \tau \}$: $\tau \in [0, 2 \pi)$ parametrizes the $S^1$, whereas $\vartheta \in [0, \pi]$ and $\varphi \in [0, 2 \pi)$ are the polar and azimuthal angles of the two-sphere (see eq.~\eqref{eq:coords_tor}).
The $S^2_\varepsilon$ surrounds the Dirac monopole at each point along its worldline, and the integral of $F$ over this $S^2_\varepsilon$ yields the enclosed magnetic charge
\begin{equation} \label{eq:intF}
    \frac{1}{2 \pi} \int_{S^2_\varepsilon} F = m \in \mathbb{Z} \ ,
\end{equation}
provided $\varepsilon$ is sufficiently small for the $S^2_\varepsilon$ and the worldline $S^1$ to link.\footnote{It follows from eq.~\eqref{eq:intF}, where the $S^2_\varepsilon$ is taken to be oriented $\textit{positively}$, that we define the magnetic charge to be positive for a positive $F_{\vartheta\phi}$ near the monopole, consistent with e.g.~ref.~\cite{Tong:2005un}. This is equivalent to defining $B_i\equiv +\frac{1}{2}\epsilon_{ijk}F_{jk}$. The opposite sign convention occasionally appears in the literature, see e.g.~refs.~\cite{Shifman:2012zz,Weinberg:2012pjx}. Working in that convention would amount to taking $m\rightarrow -m$ everywhere in this section.\label{footnote:sign_ambig}} Using Stokes' theorem, one can equivalently write the magnetic charge as an integral over the difference between the two gauge fields $A^\mathrm{(out)}$ and $A^\mathrm{(in)}$ (given by eq.~\eqref{eq:Aout} and eq.~\eqref{eq:Ain}, respectively) on the magnetic monopole equator
\begin{subequations}
\label{eq:abelian-equator-integral}
\begin{align}
    \frac{1}{2\pi}\int_{S^2_\varepsilon} F &= \frac{1}{2\pi}\left[\int_{S^2_{\varepsilon,N}} \dd A^\mathrm{(out)}-\int_{S^2_{\varepsilon,S}} \dd A^\mathrm{(in)}\right] \\ 
    &=\frac{1}{2\pi}\int_{S^1_\varepsilon} \left[A^\mathrm{(out)}-A^\mathrm{(in)}\right] \\ 
    & = \frac{1}{2\pi}\int_{S^1_\varepsilon} ig \dd g^{-1} 
    =m \, ,
\end{align}
\end{subequations}
where $g$ is the gauge transformation parameterized by eq.~\eqref{eq:gauge_phi}, $S^1_\varepsilon$ is the equator of $S^2_\varepsilon$ parameterized by $\{\vartheta=\pi/2,\,\varphi\in[0,2\pi)\}$, and $S^2_{\varepsilon,N}$ and $S^2_{\varepsilon,S}$ are the northern and southern hemispheres of $S^2_\varepsilon$, respectively. Note that we have picked $S^1_\varepsilon$ to lie within the domain of both gauge fields. This calculation shows that the integer magnetic charge of the monopole furnishes an element of the homotopy group $\pi_1(U(1))=\mathbb{Z}$ --- a fact that will become important shortly in describing the topological origin of the dyon loop instanton number.

By itself, the magnetic loop configuration we have discussed thus far does not carry instanton number. Indeed, it is obvious from eq.~\eqref{eq:Fmonopoleloop} or \eqref{eq:Fmonopoleloop_tor} that $F \wedge F$ must vanish, since $F$ lacks enough non-zero components for its wedge product with itself to be non-zero. This brings us to our main claim: a configuration corresponding to a closed monopole worldline carries non-zero instanton number provided the gauge field $A$ simultaneously winds $n\neq 0$ times around the $S^1$ of the internal boundary. That is, in addition to eq.~\eqref{eq:intF}, we also require integer quantization of the winding around the loop
\begin{equation}
    \frac{1}{2 \pi} \int_{S^1} A = n  \in \mathbb{Z} \, ,\label{eq:n_wind}
\end{equation}
consistent with a $U(1)$ holonomy with trivial phase. Having $n \neq 0$ requires a non-vanishing $A_\tau$ component on the boundary $S^1$, which we take to be pure gauge\footnote{On the boundary $S^1$, $A_\tau$ must be independent of $\vartheta$ and $\varphi$, as these are both undefined at the location of the loop (see eq.~\eqref{eq:coords_tor}). Any $\tau$ dependence may be removed via an appropriate $\tau$-dependent gauge transformation.}
\begin{equation} \label{eq:Atau_S1}
    A_\tau \Big|_{S^1} = ig\partial_\tau g^{-1} =  n \,, \qquad g = e^{in\tau}\, , \quad n \in \mathbb{Z} \ .
\end{equation}
Hence, the worldline $A_\tau$ winding furnishes another element of the homotopy group $\pi_1(U(1))$, distinct from the one associated with magnetic charge in eq.~\eqref{eq:abelian-equator-integral}.

An important caveat to this statement is that in certain UV completions, the winding need not strictly be an integer. For example, as we will see in~\sec{sec:free_dyon_su2u1} and~\sec{sec:R4-SU2-to-U1}, if the Abelian gauge field is embedded in an $SU(2)$ gauge group, the gauge field winding can be \textit{half-integer} (provided one works in a normalization where the minimal electric charge is $\pm 1$). This comes from the $\mathbb{Z}_2$ one-form center symmetry of the $SU(2)$ gauge theory, which becomes an electric $\mathbb{Z}_2$ one-form symmetry for the Abelian gauge field in the infrared. In that case, the half-integer winding is compensated by an instanton number prefactor that is a factor of two larger than one would naively expect from the bottom-up $U(1)$ theory. We will see in~\sec{sec:free_dyon_su2u1} and~\sec{sec:R4-SU2-to-U1} that this means that in such a UV completion, the Abelian instanton number is still integer valued (as opposed to half-integer valued). With this important caveat established, we keep to integer quantization conditions for the magnetic charge and worldline winding for the rest of this section.

We are now ready to compute the instanton number $I$ of the monopole loop + worldline winding field configuration. To highlight various aspects of how the topology of the field configuration yields a quantized instanton number, we do this in two ways: first by direct bulk integration of $F\wedge F$, and then via integrating the Abelian Chern-Simons current $A\wedge F$ on all singular surfaces associated with the gauge field. First, we directly evaluate the bulk $F\wedge F$ integral
\begin{subequations}
\label{eq:FwedgeF_tot}
\begin{align} 
    I\equiv \frac{1}{8\pi^2}\int_{\mathbb{R}^4} F \wedge F
    & = \frac{1}{4\pi^2}  \int_0^\infty \dd \lambda \int_0^{2 \pi} \dd \varphi \int_0^\pi \dd \vartheta  \int_0^{2 \pi} \dd \tau \, F_{\lambda \tau} F_{\varphi \vartheta} \label{eq:FwedgeF} \\[5pt]
    & = - \frac{1}{4\pi^2} \int_0^{2 \pi} \dd \varphi \int_0^\pi \dd \vartheta \int_0^{2\pi} \dd \tau \, A_\tau F_{\vartheta \varphi} \Big|_\mathrm{worldline}\label{eq:FwedgeF_Stokes} \\[5pt]
    & = - \frac{1}{4\pi^2} \left( \int_{S^1} A \right) \left( \int_{S^2_\varepsilon} F \right)\label{eq:FwedgeF_factor} \\[5pt]
    & = - n m \, ,\label{eq:FwedgeF_final}
\end{align}
\end{subequations}
where the subscript ``worldline'' indicates that the integral should be carried out in the limit $u^2+(v-R)^2\rightarrow 0$, or equivalently $\lambda\rightarrow\infty$. In going from eq.~\eqref{eq:FwedgeF} to~\eqref{eq:FwedgeF_Stokes}, we integrated by parts and made use of the fact that the integrand vanishes at spatial infinity. For continuity, $A_\tau$ must vanish at the center of the $x_3 x_4$ plane, as $\tau$ represents the corresponding planar angle. At $\vartheta = 0$, $F_{\vartheta \varphi} = 0$, as can be seen from eq.~\eqref{eq:Fmonopoleloop_tor}. As shown in eq.~\eqref{eq:FwedgeF_factor}, the result factors out as the line integral of eq.~\eqref{eq:n_wind} times the magnetic flux  of eq.~\eqref{eq:intF}. The final result of eq.~\eqref{eq:FwedgeF_final} is therefore non-zero provided both $n$ and $m$ do not vanish, and $\int F \wedge F$ is appropriately quantized in units of $8 \pi^2$ for $n, m \in \mathbb{Z}$.

Before commenting further, we show a second way of computing the instanton number using Stokes' theorem as a first step, so that the bulk $F\wedge F$ integral is converted to two $A\wedge F$ integrals surrounding gauge field singularities. To do this, we restrict ourselves to the singular gauge field configuration $A^\mathrm{(out)}$ defined by eq.~\eqref{eq:Aout}, which has a Dirac sheet at $\{u=0, \, v \leq R\}$. First, integrating $A\wedge F$ on the $S_\varepsilon^2\times S^1$ wrapping the monopole worldline, we get
\begin{subequations}
\label{eq:abelian_instanton_tube_intgral}
\begin{align}
    -\frac{1}{8\pi^2}\int_{S^2_\varepsilon\times S^1} A\wedge F &= -\frac{1}{8\pi^2}\int_0^{2\pi} \dd\tau\int_0^\pi\dd\vartheta\int_0^{2\pi} \dd\varphi\, A_\tau F_{\vartheta\varphi} \Big|_\mathrm{worldline}\\
    &=-\frac{1}{8\pi^2} \left[\int \dd\tau A_\tau\Big|_\mathrm{worldline} \right] \left[ \int\dd\varphi\, A_\varphi\Big|_\mathrm{string} \right]\\
    &=-\frac{1}{8\pi^2} \left[ \int_0^{2\pi} \dd\tau\, n \right] \left[ \int_0^{2\pi} \dd\varphi\, m \right]\\
    &=-\frac{1}{2}nm \, ,
\end{align}
\end{subequations}
where the subscript ``string'' indicates that the integral should be carried out in the limit $u\rightarrow 0$, $\vartheta\rightarrow \pi$ and the overall minus sign comes from the induced orientation of the $\dd^4x$ volume element. Notice that this integral is equivalent to eq.~\eqref{eq:FwedgeF_tot}, but a factor of two smaller. The other ``half'' comes from the surface wrapping the singular Dirac sheet. This surface is parameterized by $\Sigma_\mathrm{sheet}\equiv\{u=\varepsilon,\, \varphi\in[0,2\pi),\, v\in[0,R],\,\tau\in[0,2\pi)\}$, so that
\begin{subequations}
\label{eq:abelian_instanton_sheet_intgral}
\begin{align}
    -\frac{1}{8\pi^2}\int_{\Sigma_\mathrm{sheet}}A\wedge F &=-\frac{1}{8\pi^2}\int_0^{2\pi} \dd\varphi\int_0^R\dd v\int_0^{2\pi} \dd\tau\, A_\varphi F_{v\tau}\Big|_\mathrm{string} \\
    &=-\frac{1}{8\pi^2} \left[\int_0^{2\pi} \dd\varphi\, A_\varphi\Big|_\mathrm{string} \right] \left[\int_0^{2\pi} \dd\tau\, A_\tau\Big|_\mathrm{worldline}\right] \\ 
    &= -\frac{1}{8\pi^2} \left[\int_0^{2\pi} \dd\varphi \, m\right]\left[\int_0^{2\pi} \dd\tau \, n\right] \\
    &= -\frac{1}{2}mn \, .
\end{align}
\end{subequations}
The overall minus sign again comes from the orientation induced by the $\dd^4x$ volume form. Summed together, these two surface integrals indeed yield $\frac{1}{8\pi^2}\int F\wedge F = -nm$, in agreement with eq.~\eqref{eq:FwedgeF_tot}. Thus, we have shown that our monopole loop + worldline winding field configuration carries Abelian instanton number, quantized in units of magnetic charge times worldline winding. We emphasize this key finding here
\begin{equation}
    \mathrm{(instanton\; number)}=-\mathrm{(magnetic\;charge)}\times\mathrm{(worldline\;winding)} \, .
\end{equation}

\begin{figure}
    \includegraphics[height=0.36\textwidth]{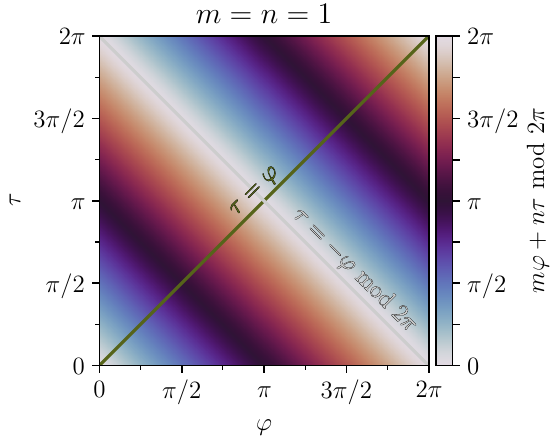}
    \includegraphics[height=0.36\textwidth]{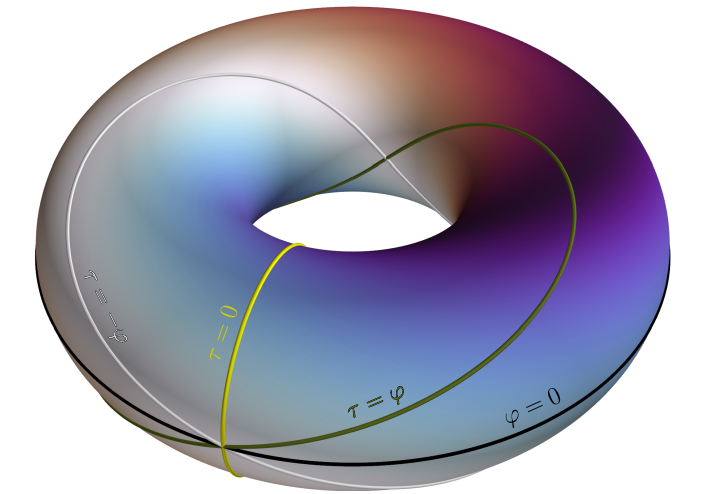}
    \caption{Visualization of (one element of) the torus homotopy group $\tau_{2}(S^1) \equiv [T^2, S^1] = \mathbb{Z} \times \mathbb{Z}$. The base space $T^2 = S^1 \times S^1$ is the product the equator of the $S^2$ surrounding the monopole of magnetic charge $m$, parametrized by $\varphi$, and of the monopole loop, parametrized by $\tau$. The yellow circle is the equator of this $S^2$ at $\tau=0$, also depicted in figure~\ref{fig:summary}. The $U(1)$ gauge field can wind around either $S^1$ independently, corresponding to magnetic charge $m$ (eq.~\eqref{eq:gauge_phi}) and winding number $n$ (eq.~\eqref{eq:n_wind}), respectively. Each homotopy group element can thus be represented by the map $(\varphi, \tau) \mapsto m \varphi + n \tau ~\mathrm{mod} ~2\pi$; only the element corresponding $m=n=1$ is shown in the figure. 
    Such maps are uniquely determined by their ``gradient'': $\alpha \mapsto (\varphi,\tau) = (m \alpha, n \alpha)$ for $\alpha \in [0,2\pi)$, tracing out the olive curve, showing isomorphism to the fundamental group of the two-torus $\pi_1(T^2)$.}
    \label{fig:homotopy}
\end{figure}

As anticipated at the start of this section, the instanton number of this magnetic worldline is entirely due to its non-trivial winding around the three-dimensional boundary $S^1 \times S^2_\varepsilon$. We visually represent the non-trivial topology of this gauge field configuration in figure~\ref{fig:homotopy}. The internal boundary $S^1 \times S^2_\varepsilon$ contains $T^2 = S^1 \times S^1_\varepsilon$ as a subspace, with the first $S^1$ being the worldline extent and the second referring to the equator of $S^2_\varepsilon$. The gauge field winding along the first $S^1$ (the worldline) furnishes an element of $\pi_1(S^1)=\mathbb{Z}$, with winding number labeled by the integer $n$, corresponding to the pure gauge integral along the monopole worldline, as shown via eq.~\eqref{eq:Atau_S1}. The pure gauge winding along $S^1_\varepsilon$ (the monopole equator) similarly furnishes an element of $\pi_1(S^1)=\mathbb{Z}$, this time labeled by the integer $m$, corresponding to the monopole magnetic charge, as shown via eq.~\eqref{eq:abelian-equator-integral}. Since the instanton number is equivalent to a product of integrals over these two circles, it furnishes an element of the torus homotopy group~\cite{fox1945torus,fox1948homotopy} $\tau_{2}(S^1) \equiv [T^2, S^1] = \mathbb{Z} \times \mathbb{Z}$. Maps classified by this homotopy group are topologically non-trivial, and can be represented by $f_{m,n}: T^2 \to U(1)$ with $f_{m,n}: (\varphi, \tau) \mapsto e^{i(m\varphi + n \tau)}$.\footnote{There is a one-to-one correspondence with maps $\widetilde{f}_{m,n}: U(1) \to T^2$ defined by $\widetilde{f}_{m,n}: e^{i\alpha} \mapsto (\varphi,\tau) = \alpha (m,n)$, proving the isomorphism $\tau_2(S^1) \cong \pi_1(T^2) \equiv [S^1, T^2] = \mathbb{Z} \times \mathbb{Z}$ to the more familiar two-torus fundamental group.} Hence, the presence of the magnetic monopole loop with $m\neq0$ allows for Abelian gauge field configurations that are well-defined and wind along and around the monopole worldline. This in turn allows the configuration to have non-zero instanton number as shown in eq.~\eqref{eq:FwedgeF_tot}.

Finally, we note that while the instanton number decomposing into the product of integers $n$ and $m$ is gauge invariant, the location of the $n$ integer winding is not. One can always perform a singular gauge transformation, e.g. of the form $g=e^{-in\tau}$, to remove the winding number along the monopole worldline, but this comes at the cost of creating a $\dd\tau$ singularity at $v=0$. This means that the field configuration now has a second singular sheet that is orthogonal to the Dirac sheet. In this ``worldline unwound'' gauge, the instanton number contribution from the surface wrapping the monopole worldline vanishes, being traded for a non-trivial contribution from the surface wrapping the new $v=0$ singular sheet. However, since having non-trivial $\tau$ winding on the monopole loop itself introduces the fewest singular surfaces, we keep to this particular ``natural'' choice of gauge in the rest of this work.
\subsection{Wound Monopoles and Dyon Loops}
\label{sec:u1dyon}

We have established that closed magnetic worldlines carry non-zero instanton number provided the $U(1)$ gauge field also winds non-trivially along the worldline $S^1$. We now show that this field configuration can be described as a dyon loop wherein the dyon's electric charge is ``smeared" rather than delta-function localized on the loop.
Indeed, the Abelian gauge field configuration satisfying both eq.~\eqref{eq:intF} and \eqref{eq:n_wind} corresponds to a solution of Maxwell's equations only in the presence of both magnetic and electric currents. The magnetic current is that of eq.\,\eqref{eq:jmonopoleloop}. The existence of our Abelian instanton then hinges on the existence of a corresponding electric current with physically sensible properties (and a sensible UV completion of the monopole singularity, the subject of~\sec{sec:free-dyon} and~\sec{sec:non-abelian-loop}).  This is hardly surprising: $\int F\wedge F \propto \int \mathbf{E}\cdot \mathbf{B}$, so a field configuration with non-zero Abelian instanton number necessarily consists of $E$- and $B$-fields with a parallel component. Additionally, having finite $A_\tau$ on the monopole worldline while requiring $A_\tau\rightarrow 0$ at the center of the loop and at infinity means $A_\tau$ must have non-trivial $u$ and $v$ dependence so that $\partial_u A_\tau$ and $\partial_v A_\tau$ source an electric field profile pointing into (or away from) the monopole worldline.

Given the symmetries of our problem, the dyon electric one-form current $j_e$ can be parameterized as
\begin{subequations}
\label{eq:jeloop}
\begin{align} 
    e^2 j_e & = \rho_e(u,v) (-\sin \tau \, \dd x_3 + \cos \tau \, \dd x_4) \label{eq:jeloopv1} \\
        & = v \rho_e(u,v) \dd \tau \ , \label{eq:jeloopv2}
\end{align}
\end{subequations}
where $\rho_e (u,v)$ is some electric charge density whose functional dependence on $u$ and $v$ will be the focus of what follows. This current can source a non-zero $A_\tau$ component: the corresponding Maxwell equation from eq.~\eqref{eq:maxwell_inhom} can be written as
\begin{equation} \label{eq:Maxwell_tau}
    (u v)^{-1} \partial_u \left( u \partial_u A_\tau \right) + \partial_v \left( v^{-1} \partial_v A_\tau \right) = \rho_e (u,v) \ .
\end{equation}
It is instructive to investigate how the electric charge density $\rho_e (u,v)$ needs to behave in various limits, given the requirements identified in the previous subsection. As discussed around eq.~\eqref{eq:Atau_S1}, $A_\tau$ is assumed to be constant on the boundary $S^1$, given the azimuthal symmetry along the $\tau$ direction in our setup. In its neighborhood, we may then consider an ansatz of the form
\begin{equation} \label{eq:Atau_ansatz_loop}
    A_\tau (r) \simeq n \left( 1 + \frac{r}{ l_e} \right) \qquad \text{with} \qquad r \equiv \sqrt{u^2 + (v-R)^2} \ll R \ ,
\end{equation}
and where $ l_e$ is a length scale associated with the extent of the electric charge distribution.
Expanding eq.~\eqref{eq:Maxwell_tau} near $r = 0$, we find the electric charge density
\begin{equation} \label{eq:rhoe_near_u1}
    \rho_e (r)
        \simeq \frac{2}{R} \frac{\partial_r A_\tau}{r}
        \simeq \frac{2 n}{r  l_e R}
        \qquad \text{for} \qquad r \ll R \ .
\end{equation}
Thus, electric charge is localized on the monopole worldline, but finiteness of $A_\tau$ on the $S^1$ restricts $\rho_e$ to diverge at most as $r^{-1}$ as $r \rightarrow 0$. This is in contrast with the magnetic current of eq.\,\eqref{eq:jmonopoleloop}, which we take to be point-like. In this sense, the dyonic field configuration that resolves the Abelian instanton identified in \sec{sec:u1monopole} does not correspond to a point-like elementary dyon. Instead, its electric charge \emph{must} be distributed over some finite length scale $ l_e$, which cannot be computed within the context of the Abelian theory in the IR. As we will see in~\sec{sec:julia-zee-dyon}, this $r^{-1}$ scaling of the electric charge density near the dyon core is precisely the behavior seen in the UV-complete Julia-Zee dyon that is present in the Georgi-Glashow model, where $\rho_e (r) \propto r^{-1}$.

An additional requirement for both $A$ and $j_e$ to be well defined is that their $\tau$-components must vanish 
as $v\rightarrow 0$, for all values of $u$. Let us consider a polynomial ansatz for $A_\tau$, as follows
\begin{equation} \label{eq:Atau_ans_v0}
    A_\tau (u,v) \simeq n C(u) \left( \frac{v}{R} \right)^p \qquad \text{for} \qquad v \ll R \ ,
\end{equation}
where $p \geq 1$ is a positive integer and $C(u)$ is some (dimensionless) function of $u$.
Plugging eq.~\eqref{eq:Atau_ans_v0} into eq.~\eqref{eq:Maxwell_tau}, we find that the behavior of the electric charge density in the vicinity of $v=0$ must be of the form
\begin{equation}
    \rho_e (u,v)
    \simeq \partial_v \left( v^{-1} \partial_v A_\tau \right)
    \simeq \frac{n}{R^3} p (p-2) C(u) \left( \frac{v}{R} \right)^{p-3} \ .
\end{equation}
$p=1$ leads to $\rho_e \propto 1/v^2$, which is incompatible with vanishing $j_{e,\tau}$ in the limit $v \rightarrow 0$ (recall eq.~\eqref{eq:jeloopv2}). For $p=2$, the leading contribution to $\rho_e$ vanishes and, instead, we have
\begin{equation} \label{eq:Atau_rhoe_v0}
    A_\tau (u,v) \simeq n C(u) \left( \frac{v}{R} \right)^2 \qquad \text{and} \qquad \rho_e (u, v) \simeq \mathcal{O} \left( \frac{v}{R} \right) \ .
\end{equation}
Eq.~\eqref{eq:Atau_rhoe_v0} is then compatible with our requirement that $A_\tau, j_{e,\tau} \rightarrow 0$ as $v\rightarrow 0$.

So far, our analysis has revealed that it is possible to introduce an electric current $j_e$ with reasonable physical properties that would source a non-singular $A_\tau$ satisfying the necessary requirements -- namely eq.~\eqref{eq:n_wind}.
An additional consideration concerns the asymptotic behavior of the gauge field for this Abelian instanton.
Working in Lorentz gauge (in keeping with our discussion in section~\ref{sec:u1monopole}), we can readily evaluate the asymptotic form of $A_\tau$ using the appropriate 4-dimensional Green's function, as follows
\begin{align}
    A_\tau (x)
    = e^2 \int \dd^4 x' \frac{x_3 j_{e,4} (x') - x_4 j_{e,3} (x')}{4 \pi^2 (x-x')^2}
    \simeq \frac{q R^2 v^2}{(u^2 + v^2)^2} + \dots \quad \text{for} \quad x^2 = u^2 + v^2 \gg R^2 \, \label{eq:Amumultiple}
\end{align}
where in the last step we have only made explicit the leading (dipole) term in a multipole expansion. The dipole coefficient $qR^2$ appearing in eq.~\eqref{eq:Amumultiple} is given by
\begin{equation} \label{eq:dipolecoeff}
    q R^2 = - \iint \dd u \, \dd v \, u v^2 \rho_e (u,v) \ ,
\end{equation}
where it is left implicit that the integrals are being performed over the entire domain of the relevant coordinates, i.e.~$u, v \in [ 0, \infty)$.

At this stage, it is helpful to look at the asymptotic behavior of the electromagnetic field strength. Far away from the loop, the Cartesian components of $F$ can be written as
\begin{align} \label{eq:Ffaraway}
\begin{split}
    F_{\mu \nu} =
    & - \frac{(m - q)}{2} \frac{4 R^2}{(x^2)^2} \left\{ \bar \eta_{3 \mu \sigma} \left( \frac{x_\nu x_\sigma}{x^2} - \frac{\delta_{\nu \sigma}}{4} \right) - (\mu \leftrightarrow \nu) \right\} \\
    & - \frac{(m + q)}{2} \frac{4 R^2}{(x^2)^2} \left\{ \eta_{3 \mu \sigma} \left( \frac{x_\nu x_\sigma}{x^2} - \frac{\delta_{\nu \sigma}}{4} \right) - (\mu \leftrightarrow \nu) \right\} + \dots \ ,
\end{split}
\end{align}
where the $\eta_{a\mu\nu}$ and $\bar{\eta}_{a\mu\nu}$ are the 't Hooft symbols~\cite[Eqs.~A3--A4]{tHooft:1976snw}, and where the dots indicate terms that are higher order in $x^2 / R^2 \ll 1$. The various terms in eq.~\eqref{eq:Ffaraway} are easily recognizable: the expressions multiplying the factors $- (m \mp q) / 2$ are precisely the third isospin components of the BPST instanton and anti-instanton respectively, in singular gauge, after making the identification $8 \rho_i^2 \rightarrow 4 R^2$ and $g \rightarrow e$, where $\rho_i$ refers to the instanton radius and $g$ to the non-Abelian gauge coupling. In this context, these are better thought of as the Abelian components of the ``constrained instantons" of the Georgi-Glashow model, in a gauge where the adjoint Higgs points entirely along the third isospin direction~\cite{Affleck:1980mp}.

In light of eq.~\eqref{eq:dipolecoeff}, it is clear that we must have $q \propto n$, since $A_\tau$ and $j_e$ must vanish for zero winding. For the field configurations identified in the preceding paragraph, further demanding that they are asymptotically (anti-)self-dual requires that $q = \mp m \in \mathbb{Z}$. Since $q \propto n$ and $n \in \mathbb{Z}$ itself, it follows that $q = n$ is the simplest option. The instanton number $I$ for these (anti-)self-dual field configurations is then given by (recall eq.~\eqref{eq:FwedgeF_tot})
\begin{equation}
    I = \frac{1}{8 \pi^2} \int F \wedge F = 
    \begin{cases}   +1 & \text{for} \ m n = - 1 \\
                    -1 & \text{for} \ m n = + 1
    \end{cases} \, ,
\end{equation}
showing that these dyon configurations indeed yield the instanton number established in~\ref{sec:u1monopole}.

\subsection{Instanton Action}
\label{sec:u1action}

Since our Abelian dyon field configuration consists of a point magnetic monopole and a smooth electric charge distribution, it is clear that a UV-complete action is needed to describe the detailed dynamics at the core of the dyon.\footnote{In fact, since our configuration simultaneously possesses electric and magnetic charge, no Abelian Lagrangian that is simultaneously local and Lorentz invariant can be written down.} This means that within the $U(1)$ gauge theory in the IR, any calculation of the action associated to our Abelian instanton is necessarily an estimate that may only be accurate within a certain parametric regime beyond which the effective theory becomes invalid.
To perform this estimate, we closely follow the approach of ref.~\cite{Affleck:1981ag} in calculating the tunneling rate for a homogeneous magnetic field to decay into pairs of magnetic monopoles. Namely, we will assume the existence of a four-dimensional UV completion featuring magnetic as well as dyonic defects, such that the dyon loop configuration described in~\sec{sec:u1monopole} and~\sec{sec:u1dyon} is resolved in terms of some unspecified underlying degrees of freedom not present at low energies --- an expectation we will see is correct in~\sec{sec:free-dyon} and~\sec{sec:non-abelian-loop} when we go to a UV-complete theory.

The corresponding Euclidean action can then schematically be written as
\begin{subequations}
\begin{align}
    S_E & = \int \dd^4 x \, \mathcal{L}_\text{UV} \label{eq:U1UVaction} \\
    & = \underbrace{ \int \dd^4 x \left\lbrace \mathcal{L}_\text{UV} - \frac{1}{4e^2} F_{\mu \nu} F_{\mu \nu} \right\rbrace }_{\Delta S_\text{UV}} + \underbrace{ \int \dd^4 x \left\lbrace  \frac{1}{4e^2} F_{\mu \nu} F_{\mu \nu} \right\rbrace }_{\Delta S_\text{IR}} \, , \label{eq:U1action_split}
\end{align}
\end{subequations}
where $\mathcal{L}_\text{UV}$ refers to the Lagrangian density of a UV completion that contains magnetic and dyonic defects, and $F_{\mu \nu}$ and $e$ refer to the field strength and gauge coupling of the $U(1)$ gauge theory in the IR.
Eq.~\eqref{eq:U1action_split} is a rearrangement of eq.~\eqref{eq:U1UVaction} after subtracting and adding the kinetic term of the Abelian gauge theory. In the following, we will describe the form of the two contributions $\Delta S_\text{UV}$ and $\Delta S_\text{IR}$, while relegating the details to appendix~\ref{sec:app_action}.

Let us first comment on the second term in eq.~\eqref{eq:U1action_split}. This is just the kinetic term of the $U(1)$ gauge theory, evaluated on our dyon loop configuration. On account of the vanishing radius of the Dirac monopole ($\varepsilon \rightarrow 0$, in the language of \ref{sec:u1monopole}), $\Delta S_\text{IR}$ diverges when evaluated on our Abelian instanton. This divergence may be regulated by introducing a smooth magnetic charge density, analogous to the electric charge density $\rho_e$ of eq.~\eqref{eq:jeloop}. As described in appendix~\ref{sec:app_action}, the resultant action takes the form
\begin{equation} \label{eq:DeltaSIR}
    \Delta S_\text{IR} = 2 \pi R \Lambda - \frac{\pi^2 (m^2 + q^2)}{e^2} + \dots \ ,
\end{equation}
where $\Lambda$ refers to the Coulomb self-energy of the (regulated) dyon loop configuration (see eq.~\eqref{eq:Lambda_def}). In the limit of vanishing monopole radius, $\Lambda \rightarrow \infty$ and the terms denoted by dots vanish.

We now turn to the first term in eq.~\eqref{eq:U1action_split}. To estimate the size of $\Delta S_\text{UV}$, we will make the assumption that the UV-complete description of our dyon loop only differs from the purely Abelian description within a distance $l_{e,m} \ll R$ from the monopole worldline. As described in appendix~\ref{sec:app_action}, one finds
\begin{equation} \label{eq:DeltaSUV}
    \Delta S_\text{UV} = 2 \pi R ( M - \Lambda ) + \dots \ ,
\end{equation}
with $M$ the dyon mass in the full UV-theory, and where the dots denote terms that vanish in the limit of zero thickness. Unlike eq.~\eqref{eq:DeltaSIR}, $\Delta S_\text{UV}$ does not contain terms of $\mathcal{O}(R^0)$.

In total, adding eq.~\eqref{eq:DeltaSIR} and \eqref{eq:DeltaSUV}, we then have
\begin{equation} \label{eq:u1SE}
    S_E = 2 \pi R M - \frac{\pi^2 (m^2 + q^2)}{e^2} + \mathcal{O} \left(l^2 M / R \right) \ ,
\end{equation}
where $l \sim l_m\sim l_e$ refers to the dyon thickness in the UV theory, which we can expect to be $l \sim M^{-1}$. An important comment is in order at this point. Obtaining the estimate in eq.~\eqref{eq:u1SE} crucially relies on the assumption that the electric and magnetic charge densities of our dyon loop configuration are localized within a small distance compared to the loop radius, that is $R M \propto R / l \gg 1$. Since eq.~\eqref{eq:u1SE} is clearly decreased by taking $R$ small, calculating the action of the instanton field configuration that dominates in the path integral is therefore not possible within the infrared effective theory alone. As $R$ becomes small, corrections to eq.~\eqref{eq:u1SE} become important, and their size depends on features that are specific to the UV completion. For example, we will see in~\sec{sec:non-abelian-loop} that in a Georgi-Glashow UV completion, the exact BPST instanton action is recovered in the $R\rightarrow 0$ limit, where the dyon loop description of the instanton breaks down.

\section{Dyons and Instantons in the Georgi-Glashow Model in $\mathbb{R}^3 \times S^1$}\label{sec:free-dyon}

We now turn to Euclidean dyon solutions in the simplest of UV completions: the Georgi-Glashow model of a non-Abelian $SU(2)$ gauge theory spontaneously broken down to a $U(1)$ gauge group by the vacuum expectation value of an adjoint (triplet) Higgs field. Its Euclidean action is given by
\begin{equation}
    S = \int \dd^4 x \, \Big\lbrace
    \frac{1}{4g^2} G^{a}_{\mu \nu} G^{a}_{\mu \nu}
    +\frac{1}{2} D_\mu \Phi^a D_\mu \Phi^a  
    +\frac{\lambda}{4} \left(\Phi^a \Phi^a - \Phi_0^2 \right)^2\Big\rbrace \, , \label{eq:georgi_glashow_action}
\end{equation}
where the $SU(2)$ field strength is given by $G_{\mu\nu}^a=\partial_\mu A^a_\nu -\partial_\nu A_\mu^a + \epsilon^{abc}A_\mu^b A_\nu^c$, the Higgs covariant derivative is $D_\mu \Phi^a \equiv \partial_\mu \Phi^a + \varepsilon^{a b c} A^b_\mu \Phi^c$, and where $\Phi_0$ is a mass-dimension-one constant (the vacuum expectation value of the adjoint Higgs). For simplicity, we keep to a normalization where the gauge coupling $g^2$ is divided out in front of the kinetic term. Before constructing a closed dyon loop in this theory in Euclidean $\mathbb{R}^4$ in~\sec{sec:non-abelian-loop}, we first study its simpler $\mathbb{R}^3\times S^1$ Euclidean space counterpart. We do this to (i) illustrate that the Abelian instanton nature of these dyon solutions already occurs in this simpler spacetime, and (ii) to get a solid analytic handle on closed dyon worldline field configuration, which we will use in~\sec{sec:non-abelian-loop} to analyze the more complicated $\mathbb{R}^4$-embedded loop.

We begin in section~\ref{sec:julia-zee-dyon} with a review of the Julia-Zee dyon~\cite{Julia:1975ff} --- an extension of the familiar 't Hooft-Polyakov monopole that carries both magnetic and electric charge. As was first discussed by refs.~\cite{Marciano:1976as, Christ:1979iw}, this field configuration carries non-zero (and finite) instanton number on $\mathbb{R}^3 \times S^1$. In~\sec{sec:free_dyon_su2u1}, we study the Julia-Zee dyon in singular gauge where the Abelian and $W$ gauge field subcomponents can be fully separated from one another and show that the instanton number can be calculated entirely from the Abelian subcomponent, making clear that the dyon can sensibly be interpreted as an Abelian instanton if one does not have detailed knowledge of the $W$ and Higgs fields.

\subsection{The Julia-Zee Dyon}\label{sec:julia-zee-dyon}

The Julia-Zee dyon is a generalization of the familiar 't Hooft-Polyakov monopole that carries both magnetic and electric charge. It is usually stated as a solution to the Lorentzian Georgi-Glashow equations of motion, but an equivalent construction exists in Euclidean space. Since we are ultimately interested in Euclidean dyon loops, we work with a Euclidean Julia-Zee field configuration in what follows. In so-called hedgehog or static gauge, this configuration is given by
\begin{equation} \label{eq:JZ_dyon}
    \Phi^a = \hat r^a \Phi_0 h(r), \qquad A^a_i =\varepsilon_{aij} \hat r^j \frac{k(r)}{r} \qquad \text{and} \qquad A^a_4 = \hat r^a j(r) \ ,
\end{equation}
where $h$, $k$ and $j$ are smooth functions of the spatial radial coordinate $r \equiv \sqrt{x_1^2 + x_2^2 + x_3^2}$ so that the dyon worldline extends along the $x_4$ direction. The functions take on limiting values
\begin{equation} \label{eq:JZ_dyon_bc}
    h(r), k(r), j(r) \xrightarrow{r \rightarrow 0} 0
    \qquad \text{and} \qquad h(r), k(r) \xrightarrow{r \rightarrow \infty} 1 \ ,
\end{equation}
whereas
\begin{equation} \label{eq:jfar}
    j(r) \simeq \mu  + \frac{g Q_e}{4 \pi r} + \mathcal{O} \left( \frac{1}{r^2} \right) \quad \text{as} \quad r \rightarrow \infty \ .
\end{equation}
Above, $\mu$ is a constant with mass dimension equal to unity and is related to the energy associated with exciting a 't Hooft-Polyakov monopole into a dyon, and $Q_e$ refers to the electric charge of the dyon under the unbroken $U(1)$.
The $\Phi^a$ and $A^a_i$ components in eq.~\eqref{eq:JZ_dyon} are precisely those of the 't Hooft-Polyakov monopole, whereas the non-trivial profile of the $A^a_4$ components endows the monopole with non-zero electric charge.
Asymptotically, the Abelian magnetic and electric field components read\footnote{Here, the sign of $\mathcal{B}_i$ has been picked to be consistent with~\sec{sec:abelian-loop}. See footnote~\ref{footnote:sign_ambig} for a brief discussion of alternate conventions.}
\begin{equation} \label{eq:EandB_u1}
    \mathcal{B}_i
    \equiv  \frac{1}{2 g} \varepsilon_{ijk} \mathcal{F}_{jk}
    \simeq -\frac{\hat r^i}{g r^2}
    \qquad \text{and} \qquad
    \mathcal{E}_i
    \equiv - \frac{1}{g} \mathcal{F}_{i 4} 
    \simeq \hat r^i \frac{Q_e}{4 \pi r^2} \quad \text{as} \quad r \rightarrow \infty \ . 
\end{equation}
Here, $\mathcal{F}_{\mu \nu}$ is the gauge invariant 't Hooft tensor~\cite{tHooft:1974kcl}, corresponding to the field strength of the unbroken $U(1)$ component, given by
\begin{equation} \label{eq:thooft_tensor}
    \mathcal{F}_{\mu\nu} \equiv \frac{1}{|\Phi|} \Phi^a G^a_{\mu\nu} - \frac{1}{|\Phi|^3} \epsilon^{abc} \Phi^a D_\mu \Phi^b D_\nu \Phi^c \ ,
\end{equation}
where $|\Phi| \equiv \sqrt{\Phi^a \Phi^a} = \Phi_0 |h(r)|$. In hedgehog gauge, the electric field is sourced by the first term, while the magnetic field is sourced by the second. The second term in eq.~\eqref{eq:thooft_tensor} is what allows $\mathcal{F}_{\mu\nu}$ to violate the standard $U(1)$ Bianchi identity even though $A_\mu^a$ does not exhibit singular behavior. Note in particular that in our convention, the outward pointing hedgehog configuration carries \textit{negative} magnetic charge, matching the ``topological'' definition of the $B$-field from~\sec{sec:abelian-loop}.

We now establish the relation between the parameter $\mu$ and the electric charge $Q_e$ of the field configuration. Using Gauss' law, we can equate the bulk charge integral with a surface integral at infinity, yielding the relation
\begin{equation}
    Q_e =4\pi \int_0^\infty \dd r\,  \rho_e(r)=-\frac{8 \pi}{g} \int_0^\infty \dd r \, [1-k(r)]^2 j(r) =  \int_{S^2_\infty} \dd \mathbf{S}\cdot\bm{\mathcal{E}} \sim - \frac{4 \pi}{g} \frac{\mu}{m_W} \ ,
\end{equation}
where $m_W=g \Phi_0$ is the $W$ boson mass and $\dd \mathbf{S}$ is the outward oriented vector area element of $S^2_\infty$. The electric charge profile of the dyon $\rho_e (r) = - \frac{2}{g r^2} [1 - k(r)]^2 j(r)$ is the $U(1)$ Noether charge associated with the charged $W$ field~\cite{Weinberg:2012pjx}. Exact equality is achieved in the BPS limit, defined by $\lambda \to 0$.
We can also relate $\mu$ and $Q_e$ to the extra mass $M_\Delta$ added to the 't Hooft-Polyakov monopole due to the additional dyonic electric charge. This mass is given by
\begin{equation}
    M_\Delta=\frac{1}{2 g^2} \int \dd^3 {\bf x} \, G_{i4}^a G_{i4}^a = - \frac{\mu Q_e}{2 g} \sim m_W \frac{Q_e^2}{4 \pi} \ .
\end{equation}
where exact equality again is achieved in the BPS limit.

Next, we discuss the instanton number associated with this field configuration. It was first noted in \cite{Marciano:1976as,Christ:1979iw} that the integral of $\text{tr} \left[ G \wedge G\right]$ for the Julia-Zee dyon is non-zero and finite when computed over a finite time interval.\footnote{This was noted in a \textit{Lorentzian} spacetime, but holds in Euclidean signature as well.} Specifically, the integral over the spatial directions yields
\begin{equation}
    \frac{1}{8 \pi^2} \int_{\mathbb{R}^3} \dd^3 x \, \text{tr} \left[ G \wedge G\right] =  -\frac{\mu}{2 \pi} \ .
\end{equation}
Further integrating over a periodic time direction of length $T$ (the circumference of the $S^1$ parameterized by $x_4$), one finds
\begin{align}
    I = \frac{1}{8 \pi^2} \int_{{\mathbb{R}^3} \times S^1} \dd^4 x \, \text{tr} \left[ G \wedge G\right]
    & =  -\frac{\mu T }{2 \pi} \label{eq:GG_JZ} = -n \, \textrm{sign}\, \mu \ \in \ \mathbb{Z} \quad \text{for} \quad T = \frac{2 \pi n}{|\mu|} \ ,
\end{align}
where $n \in \mathbb{Z}$. Provided that $T$ is an integer multiple of $2 \pi / |\mu|$, the instanton number of the Julia-Zee dyon is quantized in units of $8\pi^2$. Classically, other intervals are allowed, but this quantization of $T$ is required for the dyon to have holonomy $\pm 1$.\footnote{In a time-dependent gauge, the quantization of $T$ can equivalently be understood as requiring $A_\mu^a$ to undergo a full $U(1)$ rotation at infinity, as shown in ref.~\cite{Shnir:2005vvi}.} This also means that a properly quantized infinite worldline carries infinite instanton number. Classically, however, one can take the limit $\mu \rightarrow 0$ while simultaneously taking $T \propto 1 / |\mu| \rightarrow \infty$ and recover a pure 't Hooft-Polyakov (with no electric charge) that carries instanton number unity~\cite{Christ:1979iw}.
\subsection{Abelian Reduction} \label{sec:free_dyon_su2u1}

We now connect the free Julia-Zee dyon with its pure $U(1)$ remnant, and show that in a gauge where the symmetry broken gauge field components are projected out, the instanton number is effectively carried by the Abelian degrees of freedom. To do this, we perform a singular gauge transformation on the Julia-Zee field configuration of eq.~\eqref{eq:JZ_dyon} characterized by the following group element
\begin{equation}
    \Omega \equiv \Omega_2 \Omega_1 \quad \text{with} \quad \Omega_2 \equiv e^{- \frac{i}{2} \mu x_4 \sigma^3} \quad \text{and} \quad \Omega_1 \equiv e^{-\frac{i}{2} \varphi \sigma_3}e^{\frac{i}{2} \theta \sigma_2}e^{\frac{i}{2}\varphi \sigma_3} \ ,
\end{equation}
with the non-Abelian gauge field $A_\mu$ transforming as $A_\mu\rightarrow \Omega A_\mu\Omega^\dagger+i\Omega \partial_\mu \Omega^\dagger$. The gauge transformation generated by $\Omega_1$ transforms the Julia-Zee dyon of eq.~\eqref{eq:JZ_dyon} from hedgehog to string gauge in which $\hat{\Phi}^a=\delta^{a3}$ everywhere in space, introducing a Dirac string singularity along $\theta = \pi$. The second gauge transformation generated by $\Omega_2$ brings the non-trivial behavior of $A^a_4$ from spatial infinity to the location of the dyon.
In this gauge, we can identify the Abelian gauge field subcomponent $\mathcal{A}$ and the $W$-boson subcomponent of the $SU(2)$ gauge field as
\begin{equation}
    \mathcal{A} \equiv \frac{1}{2} A^3 \qquad \text{and} \qquad W \equiv \frac{1}{\sqrt{2}} \left( A^1 + i A^2 \right) \ , \label{eq:singular_separation}
\end{equation}
with the relative factor of $2$ between $\mathcal{A}$ and $A^3$ such that $W$ carries $2$ units of electric charge under the unbroken $U(1)$ factor.\footnote{The alternate definition $\mathcal{A}\equiv A^3$ occasionally appears in the literature (see e.g.~ref.~\cite{Weinberg:2012pjx}). In this other convention, the $W$ carries electric charge $1$, but the minimal electric charge unit is $1/2$ (and correspondingly, magnetic flux comes in integer multiples of $4\pi$).} The Euclidean Georgi-Glashow action now takes the form~\cite{Weinberg:2012pjx}
\begin{multline}
    S = \int \dd^4x\Big\lbrace \frac{1}{g^2}\left(\mathcal{F}_{\mu\nu}-M_{\mu\nu}\right)^2+\frac{1}{2g^2}|D_\mu W_\nu-D_\nu W_\mu|^2\\
    +\frac{1}{2}(\partial_\mu\Phi)^2+\Phi^2|W_\mu|^2+\frac{\lambda}{4}\left(\Phi^2-\Phi_0^2\right)^2\Big\rbrace \, ,
\end{multline}
where $\Phi\equiv\Phi^3$ and $D_\mu W_\nu =\partial_\mu W_\nu + 2i\mathcal{A}_\mu W_\nu$, reflecting the charge of the $W$ boson. $M_{\mu\nu}$ is an electromagnetic moment defined by $M_{\mu\nu}=\frac{i}{2}(W^*_\mu W_\nu-W_\nu^*W_\mu)$. It serves to cancel the new Dirac monopole singularities in $\mathcal{F}_{\mu\nu}$, keeping the total action finite. The various Julia-Zee dyon field components are now given by
\begin{gather} \label{eq:phi_A_singular}
    \Phi^a = \Phi_0 h(r) \delta^{a3} \ ,  \quad
    \mathcal{A} =  \frac{1}{2} (\cos\theta - 1) \dd \varphi +  \frac{1}{2} \mathcal{J} (r) \dd x_4 \quad \text{with} \quad \ \mathcal{J} (r) \equiv j(r) - \mu \ ,
\end{gather}
as well as
\begin{equation} \label{eq:W_singular}
    W = \frac{1}{\sqrt{2}} \left[ 1 - k(r) \right] e^{i (\varphi + \mu x_4)} \left( i \, \dd \theta - \sin \theta \, \dd \varphi \right) \ .
\end{equation}
Notice that the asymptotic boundary conditions for the various functions in eq.~\eqref{eq:JZ_dyon_bc} now translate into
\begin{equation} \label{eq:JZ_bc_singular}
    \mathcal{J}(r), \left[1 - k(r)\right] \xrightarrow{r \rightarrow \infty} 0 \ , \quad \left[1 - k(r)\right] \xrightarrow{r \rightarrow 0} 1
    \quad \text{and} \quad \mathcal{J}(r) \xrightarrow{r \rightarrow 0} - \mu \ .
\end{equation}

In the gauge defined by eqs.~\eqref{eq:phi_A_singular}--\eqref{eq:W_singular}, the Abelian 't Hooft tensor is entirely aligned along the third isospin direction, i.e.~
\begin{align}
    \mathcal{F} & = \dd \mathcal{A} = \frac{1}{2}G^3 -  \frac{1}{2}i W \wedge W^* \ .
\end{align}
And the instanton number simplifies to
\begin{subequations} \label{eq:I_string_gauge_tot}
\begin{align}\label{eq:string-gauge-instanton-number}
    \frac{1}{8 \pi^2} \int_{\mathbb{R}^3 \times S^1} \text{tr} (G \wedge G ) & = \frac{1}{4\pi^2}\left[\int_{\mathbb{R}^3\times S^1}\left(\mathcal{F}-M\right)\wedge \left(\mathcal{F}-M\right) + \frac{1}{2}DW\wedge DW^*\right]\, ,\\[5pt]
    &= \frac{1}{4 \pi^2} \int_{\mathbb{R}^3 \times S^1} \mathcal{F} \wedge \mathcal{F}
    +  \frac{1}{8 \pi^2} \int_{\mathbb{R}^3 \times S^1} \dd ( W \wedge D W^* ) \label{eq:I_string_gauge} \, ,
\end{align}
\end{subequations}
where we have used the fact that the $M\wedge M$ term vanishes by antisymmetry and that the mixed $\mathcal{F}\wedge M$ terms cancel with the $i\mathcal{A}\wedge W \wedge D W^*$ terms after integrating by parts. Eq.~\eqref{eq:string-gauge-instanton-number} holds generally in string gauge. We now examine how the instanton number integral simplifies for our Euclidean Julia-Zee dyon configuration. To evaluate eq.~\eqref{eq:I_string_gauge} for the Julia-Zee dyon, we employ the second procedure described in section~\ref{sec:u1monopole}: we use Stokes' theorem to integrate over all boundaries present in the Abelian gauge field of eq.~\eqref{eq:phi_A_singular} and the $W$ gauge field of eq.~\eqref{eq:W_singular}. For simplicity, we restrict our calculation to period $T=2\pi /|\mu|$ and $\mathrm{sign}\,\mu = -1$. corresponding to a dyon with negative magnetic charge and positive electric charge, but this can easily be generalized.

First, the $W$ gauge field contribution. This field is singular on the dyon loop itself, but smooth everywhere else. Hence, Stokes' theorem picks up a $-S_\varepsilon^2\times S^1$ surface wrapping the dyon worldline (with the minus sign indicating that the $S^2_\varepsilon$ surface is oriented with an inward pointing normal vector) and an $S_\infty^2\times S^1$ surface at the boundary of the Euclidean space, yielding
\begin{equation}
    \frac{1}{8\pi^2}\int_{\mathbb{R}^3\times S^1}\dd\left(W\wedge D W^*\right) = \frac{1}{8\pi^2} \int_{\left(S^2_\infty-S^2_\varepsilon\right)\times S^1} W\wedge DW^* \, ,
\end{equation}
but the \textit{integrand} vanishes completely on both surfaces
\begin{equation}
    (W\wedge D W^*)_{\theta\varphi x_4}\Big|_{r=0}^{r=\infty}=\sin\theta\left[1-k(r)\right]^2\left[\mu+\mathcal{J}(r)\right]\Big|_{r=0}^{r=\infty}=0 \, ,
\end{equation}
where the last equality follows from the boundary conditions in eq.~\eqref{eq:JZ_bc_singular}. Due to the vanishing of this $W$-boson contribution, it is clear that the instanton number must come solely from the Abelian $\mathcal{F}\wedge\mathcal{F}$ term. We now evaluate that contribution explicitly. Using Stokes' theorem, we decompose the bulk $\mathcal{F}\wedge{\mathcal{F}}$ integral into three surface integrals: two on the $S^2_\varepsilon\times S^1$ and $S^2_\infty\times S^1$ surfaces surrounding the dyon worldline in its immediate neighborhood and at infinity, respectively, and one on the surface wrapping the Dirac sheet. This latter surface is parameterized by $\Sigma_\mathrm{sheet}\equiv\{\theta=\pi-\varepsilon,\, r\in[0,\infty),\, \varphi\in[0,2\pi),\,x_4\in[0,T)\}$. In analogy to eq.~\eqref{eq:abelian_instanton_tube_intgral}, the first integral yields
\begin{subequations}
\begin{align}
    \frac{1}{4\pi^2}\int_\mathrm{\left(S^2_\infty-S^2_\varepsilon\right)\times S^1} \mathcal{A}\wedge \mathcal{F} &= -\frac{1}{16\pi^2}\int_0^\pi\dd\theta\int_0^{2\pi}\dd\varphi\int_0^T\dd x_4 \,\mathcal{J}(r)\sin\theta\Big|_{r=0}^{r=\infty} \\ 
    &= -\frac{1}{4\pi}T \left[\mathcal{J}(\infty)-\mathcal{J}(0)\right] 
    =\frac{1}{2} \, .
\end{align}
\end{subequations}
Similarly, in analogy to eq.~\eqref{eq:abelian_instanton_sheet_intgral}, the integral along the surface wrapping the Dirac sheet yields
\begin{subequations}
\begin{align}
    -\frac{1}{4\pi^2}\int_{\Sigma_\mathrm{sheet}} \mathcal{A}\wedge \mathcal{F} &= \frac{1}{16\pi^2}\int_0^\infty\dd r\int_0^{2\pi}\dd\varphi\int_0^T\dd x_4 \,\left(\cos\theta-1\right)\mathcal{J}'(r)\Big|_{\theta=\pi}  \\ 
    & = -\frac{1}{4\pi}T\left[\mathcal{J}(\infty)-\mathcal{J}(0)\right] 
    = \frac{1}{2} \, ,
\end{align}
\end{subequations}
where we have used the fact that $\theta=\pi$ near the Dirac sheet and where the overall minus sign comes from the orientation induced by the $\dd^4x$ volume element. These surfaces are two separate contributions due to $\mathcal{A}_{x_4} \mathcal{F}_{\theta\phi}$ only being singular on the loop, and $\mathcal{A}_\varphi\mathcal{F}_{r x_4}$ being singular on the Dirac sheet. The total instanton number thus evaluates to
\begin{equation}
    I=\frac{1}{4\pi^2}\int \mathcal{F}\wedge\mathcal{F}= \frac{1}{2}+\frac{1}{2} =1\, .
\end{equation}
We have thus shown that the Julia-Zee dyon instanton number can be calculated entirely from the Abelian field strength $\mathcal{F}$, with the plus sign coming from the dyon having negative magnetic charge and positive electric charge.

Some general remarks are in order. First, we emphasize that $\mathcal{F}$, as defined via the 't Hooft tensor~\eqref{eq:thooft_tensor}, is gauge invariant --- in any gauge, the instanton number can be calculated by just integrating $\mathcal{F}\wedge\mathcal{F}$. However, the instanton number being calculable from just $\dd A^3$ only holds in string gauge, since it is only in this gauge that $\mathcal{F}\equiv\dd A^3/2$. The reason the instanton number can reside in the $A^3$ component only in string gauge is that the non-trivial gauge field behavior has been transported to the $A^3$ component, which we see in form of the singular monopole worldline and the singular Dirac sheet. Second, note that in the Georgi-Glashow model, the Abelian instanton number comes with a $1/(4\pi^2)$ prefactor, which is a factor of two larger than the simple ``bottom-up'' instanton number prefactor discussed in~\sec{sec:abelian-loop}. This factor of two has been discussed previously in ref.~\cite{Garcia-Valdecasas:2024cqn} in the context of generalized symmetry breaking. Here, this factor is exactly compensated by the dyon worldline having half-integer gauge field winding along the $S^1$ ($A_\tau = \frac{1}{2}\dd\tau$ on the loop), yielding instanton number $+1$ in spite of the doubled prefactor. This half-integer winding comes from the $\mathbb{Z}_2$ one-form center symmetry of the $SU(2)$ gauge group. More generally, the winding scales exactly with the electric charge of the Julia-Zee configuration, meaning the instanton number can be equivalently understood as the product of the integer electric and integer magnetic charge.
\section{Dyon Loops and Instantons in the Georgi-Glashow Model in $\mathbb{R}^4$}\label{sec:non-abelian-loop}
We now consider the UV-complete equivalent of the dyon loops discussed in \sec{sec:abelian-loop} using a Georgi-Glashow UV completion of the $U(1)$ gauge group. Unlike the free dyon configurations on $\mathbb{R}^3 \times S^1$ examined in \sec{sec:free-dyon}, here we study non-Abelian gauge and Higgs field configurations corresponding to closed dyon loops embedded in $\mathbb{R}^4$. Despite appearances, this is a drastically different setup from the one in  \sec{sec:free-dyon}: (i) the global topology differs --- in particular on the infinite three-sphere $S^3_\infty$ and in the region encircled by the dyon worldline, and (ii) any Euclidean time slice intersecting the loop contains a two-dyon configuration, which, due to the non-linearity of the Yang-Mills-Higgs equations, is not a simple superposition of free dyons. Although one could attempt to construct suitable configurations starting from singular gauge and applying a large gauge transformation to remove singularities~\cite{Arafune:1974uy, Bais:1976fr}, we adopt an alternative approach. We begin from a \emph{non-singular} ansatz, setting the Higgs field as a hedgehog configuration at a fixed radius $v = R$, and selecting gauge fields with BPST instanton topology at spatial infinity.
We then employ a numerical relaxation algorithm to minimize the Euclidean action at fixed instanton number: the $I=1$ sector.
This approach is advantageous for two reasons. First, we explicitly relate the dyon loop to the (Higgsed) BPST instanton, aligning with the expectation from sections~\ref{sec:abelian-loop} and~\ref{sec:free-dyon} that Julia-Zee dyon loops carry instanton number. Second, an iterative numerical relaxation algorithm performs best with fields that are regular everywhere on $\mathbb{R}^4$. After obtaining the minimal-action configurations, we switch to singular gauge, clearly separating the Abelian and $W$ degrees of freedom. This allows us to directly match the Georgi-Glashow dyon loop to the Abelian configuration detailed in \sec{sec:abelian-loop}.

Our procedure is as follows. 
We construct a gauge and Higgs field ansatz for the dyon loop in \sec{sec:ansatz}. We numerically relax these configurations to the minimum-action point subject to the constraint of a monopole defect loop at various radii $R$ in \sec{sec:relaxation}, and confirm that they reduce to the familiar (Higgsed) BPST instanton as $R \to 0$.
In \sec{sec:R4-SU2-to-U1}, we examine our solution in singular gauge, demonstrating explicitly that the instanton number is captured by the Abelian subcomponent of $\mathrm{tr}[G\wedge G]$ via an integral over the dyon worldline and its associated Dirac sheet, consistent with the description of \sec{sec:abelian-loop}.

\subsection{Georgi-Glashow Dyon Loop in $\mathbb{R}^4$}\label{sec:yang-mills-r4}
We explicitly construct a non-Abelian dyon loop configuration within the same (Georgi-Glashow) model as in \sec{sec:free-dyon}. Our goal is a numerically tractable field solution on $\mathbb{R}^4$ that simultaneously captures the topology of a closed circular dyon worldline and a BPST instanton, while remaining free from gauge or Higgs singularities.

\subsubsection{Field Configuration Ansatz}\label{sec:ansatz}
As a starting point for obtaining numerical solutions to the Euclidean Yang-Mills-Higgs equations, we construct an ansatz that respects the anticipated double azimuthal symmetry of the dyon loop, employing the polar coordinates defined in eq.~\eqref{eq:coords_dp}. To ensure our field configuration lives in the correct topological sector ($I=1$), we fix the gauge field behavior at infinity. We use (part of) the $SU(2)$ gauge freedom to fix the Higgs isospin map $\hat{\Phi}$ everywhere in $\mathbb{R}^4$. The remaining degrees of freedom --- the Higgs magnitude $|\Phi|$ and the bulk gauge field $A$ --- will be left free to vary in \sec{sec:relaxation}, constrained only by the requirement that they remain non-singular.

Specifically, we fix the $SU(2)$ gauge field configuration to have the asymptotic form
\begin{equation}\label{eq:ymh_gauge_field_topology}
    \lim_{u^2+v^2\rightarrow\infty}A^a_\mu = \tilde{A}^a_\mu[\ell] \, ,
\end{equation}
where in matrix notation
\begin{equation}
    \tilde{A}_\mu^a[\ell=\pm 1] \equiv \frac{2}{u^2+v^2}
    \begin{pmatrix}
    +v\sin{\ell\tau} & +v\cos{\ell\tau} & -u\sin{\varphi} & -u\cos{\varphi} \\
    -v \cos{\ell\tau} & +v \sin{\ell\tau} & +u\cos{\varphi} & -u\sin{\varphi} \\
    +u \sin{\varphi} & -u\cos{\varphi} &  +v\sin{\ell\tau} & -v\cos{\ell\tau}
    \end{pmatrix} \, ,
\end{equation}
with rows corresponding to the isospin index $a$ and columns to the Euclidean spatial index $\mu$. This asymptotic gauge field configuration has topological winding number $\ell = \pm 1$ for the map $S^3_\infty\rightarrow SU(2)\cong S^3$ (an element of the homotopy group $\pi_3(S^3) = \mathbb{Z}$). For $\ell=1$, it matches the standard asymptotic BPST instanton gauge field configuration in regular gauge
\begin{equation}
\tilde{A}_\mu^a[\ell=1]=2\frac{\eta_{\mu\nu}^a x_v}{x^2} \, ,
\end{equation}
with $\eta_{\mu\nu}^a$ the 't Hooft symbols~\cite{tHooft:1976snw} and $x^2 \equiv x_1^2+x_2^2+x_3^2+x_4^2$. The anti-instanton corresponds to $\ell = -1$. If no singularities appear in the interior of the Euclidean space, this choice ensures the field configuration lies in the topological sector with instanton number $I= \ell = \pm 1$, in line with our expectation that a closed dyon loop carries instanton number that scales with the dyonic charge. 

\begin{figure}
    \centering
    \includegraphics[width=0.75\linewidth]{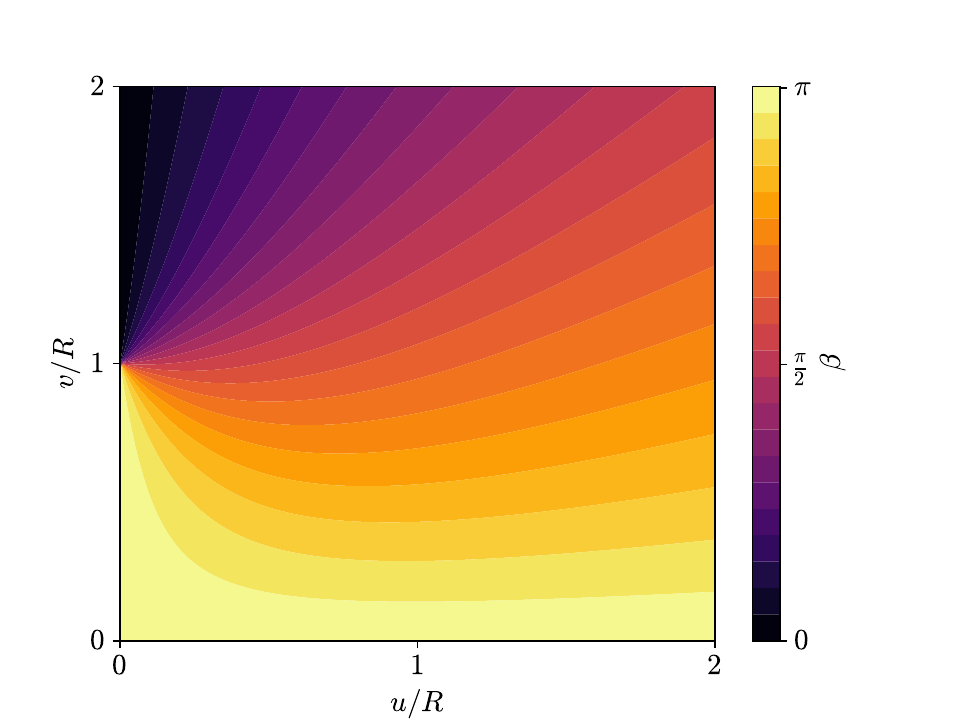}
    \caption{The $\beta$ angle defined in eq.~\eqref{eq:beta} as a function of the coordinates $u=\sqrt{x_1^2 + x_2^2}$ and $v=\sqrt{x_3^2 + x_4^2}$. This angle winds around the monopole loop in the $uv$ plane, allowing for a consistent hedgehog Higgs isospin configuration along the entire loop. Near the loop ($u\simeq 0,\, v\simeq R$), $\beta$ effectively becomes the polar angle $\vartheta_-$ of a static monopole hedgehog configuration. At spacetime infinity ($u^2+v^2\rightarrow\infty$), the angle allows the Higgs to trace a Hopf map.}
    \label{fig:beta-angle}
\end{figure}

\begin{figure}
    \centering
    \includegraphics[width=1.0\linewidth]{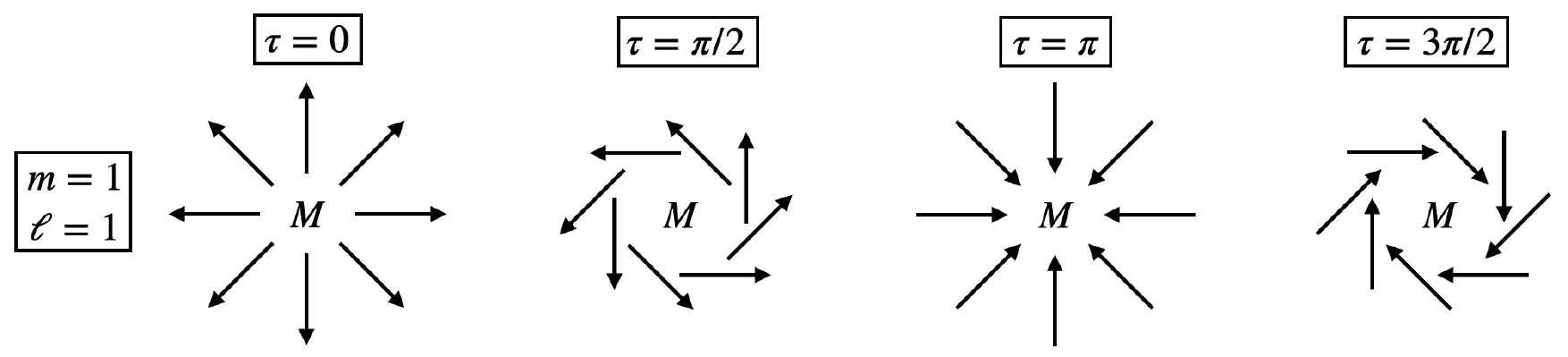}
    \caption{A visualization of the first and second component of the Higgs isospin structure from eq.~\eqref{eq:higgs_isospin} in the immediate vicinity of the dyon loop with $m=\ell=1$. While the first two isospin components rotate with $\tau$, the third remains constant. The ``twisting'' in the first two components allows for the Higgs isospin field configuration to simultaneously correspond to that of a Julia-Zee dyon in hedgehog gauge on the monopole loop, and to that of a Higgsed BPST instanton at spacetime infinity, \textit{without} the introduction of any other isospin singularities on $\mathbb{R}^4$.}
    \label{fig:hedgehog-twist-simple}
\end{figure}

Next, we fix the isospin map for the Higgs triplet
\begin{equation}\label{eq:higgs_isospin}
    \hat{\Phi}^a=   \begin{pmatrix}
        \sin \beta \cos (\varphi + \ell\tau) \\
        \sin \beta \sin (\varphi + \ell\tau) \\
        \cos \beta
        \end{pmatrix} \, ,
\end{equation}
where $\beta$ is defined as
\begin{equation}\label{eq:beta}
    \beta \equiv \vartheta_+ +\vartheta_- \, ,
\end{equation}
and $\vartheta_+$ and $\vartheta_-$ are defined via
\begin{equation}
    \tan\vartheta_+ = \frac{u}{v+R}\, , \qquad \tan\vartheta_-= \frac{u}{v-R} \, .
\end{equation}
The map of eq.~\eqref{eq:higgs_isospin} describes a hedgehog loop located at $\lbrace u=0,\, v=R\rbrace$, with $\ell$ the integer number of ``twists'' of the hedgehog around the third isospin direction over a complete traversal of the loop ($\tau = 0$ to $\tau = 2\pi$). We show a plot of $\beta$ in figure~\ref{fig:beta-angle} and a visualization of the isospin ``twist'' for $\ell=1$ in figure~\ref{fig:hedgehog-twist-simple}. Since $\hat{\Phi}^a$ is singular on the loop, regularity demands that the Higgs magnitude $|\Phi|$ vanishes there, analogous to the regular Julia-Zee dyon discussed in~\sec{sec:julia-zee-dyon}. Hence, fixing $\hat{\Phi}$ to obey eq.~\eqref{eq:higgs_isospin} amounts to manually placing down a monopole loop defect at $\lbrace u=0,\, v=R\rbrace $ with worldline parameterized by $\tau \in [0,2\pi)$. 
While the integer twist $\ell$ does not directly determine the electric charge of the dyon (as this charge is not topologically sourced), it ensures the gauge compatibility condition $(D_\mu\Phi)^a=0$ at spatial infinity, as explicitly shown in appendix~\ref{app:bpst_hopf_map}. 
Together, these choices yield a non-singular field configuration in $\mathbb{R}^4$ carrying instanton number and featuring a monopole loop defect.


We note that a Higgs map similar to eq.~\eqref{eq:higgs_isospin} was first introduced in ref.~\cite{Brower_1997}, where it appeared in the study of Yang-Mills instanton deformations within maximal Abelian projection. At spatial infinity, this map reduces to the standard Hopf map~\cite{Nakahara:2003nw}, describing a mapping $S^3_\infty\rightarrow SU(2)/U(1)=S^2$ with topological winding number $\ell \in \mathbb{Z}$.
Although one might initially suspect that this Hopf winding number always equals the instanton number or constitutes a separate invariant, this is incorrect: a singular gauge transformation can completely eliminate the non-trivial winding in the Higgs field.\footnote{One can however define a \emph{generalized Hopf invariant} that is directly related to the instanton number~\cite{Jahn:1999wx}.}

Having fixed both the asymptotic gauge field and the Higgs isospin orientation, we now specify the bulk gauge field $A_\mu^a$ and Higgs magnitude $|\Phi|$ used in our numerical relaxation algorithm. 
For the Higgs, we take
\begin{equation}
    \Phi^a = H(u,v)\hat{\Phi}^a \, , 
\end{equation}
with the radial profile function $H(u,v)$ required to satisfy
\begin{equation}
     \lim_{u^2+v^2\rightarrow\infty} H= \Phi_0 \, ,\hspace{.5in} \lim_{u\rightarrow 0,\, v\rightarrow R} H = 0 \, .
\end{equation}
Thus, $SU(2)$ symmetry is restored at the center of the monopole along $u=0, \, v=R$,  and spontaneously broken elsewhere. These conditions are necessary to keep the Euclidean action finite and to prevent singularities at the monopole core. The smooth profile $H(u,v)$ is determined numerically, with an initial guess provided in appendix~\ref{app:numerical_relaxation}.

Next, we construct the bulk gauge field configuration $A_\mu^a$. We adopt the Cho-Faddeev-Niemi gauge field decomposition~\cite{Cho:1980nx,Faddeev:1998eq}
\begin{equation}\label{eq:nonab-A}
    A_\mu^a = -K(u,v)\epsilon^{abc}\hat{\Phi}^b\partial_\mu \hat{\Phi}^c+a_\mu\hat{\Phi}^a +X_\mu^a\, ,
\end{equation}
which separates the field into three terms. The first is isospin-orthogonal to the Higgs field, with profile function $K(u,v)$; the second is isospin-aligned with the Higgs field, with the vector $a_\mu$ defined in eq.~\eqref{eq:littlea}; and the last term captures any remaining gauge field behavior --- in particular, it can be non-zero on the monopole loop defect. The profile function $K(u,v)$ is constrained to have the limiting behavior
\begin{equation}
    \lim_{u^2+v^2\rightarrow\infty}K= 1 \, ,  \hspace{.5in} \lim_{u\rightarrow 0,v\rightarrow R} K = 0 \, ,
\end{equation}
in order for $A_\mu^a$ to approach $\tilde{A}_\mu^a$ at infinity and to avoid singularities on the loop where $\hat{\Phi}^a$ is ill-defined. We choose the Higgs-aligned component $a_\mu$ to be
\begin{equation}\label{eq:littlea}
    a = \frac{2J(u,v)}{u^2+v^2+R^2}    \begin{pmatrix}
        -u\sin{\varphi}\\
        u\cos{\varphi}\\
         v\sin{\ell\tau}\\
        - v\cos{\ell \tau}
    \end{pmatrix} \, ,
\end{equation}
with the $J(u,v)$ profile function subject to the limits
\begin{equation}\label{eq:j_behavior}
    \lim_{u^2+v^2\rightarrow\infty}J= 1 \, , \hspace{15mm} \lim_{u\rightarrow 0, \, v\rightarrow R}J=0 \, ,
\end{equation}
again for $A_\mu^a$ to approach $\tilde{A}_\mu^a$ at infinity and for regularity on the monopole loop. The interpolating forms of $K(u,v)$ and $J(u,v)$ are determined through numerical relaxation, with initial guesses provided in appendix~\ref{app:numerical_relaxation}. 
Finally, the component $X_\mu^a$ is required only to vanish rapidly at infinity:
\begin{equation}\label{eq:X_behavior}
    \lim_{u^2+v^2\rightarrow\infty} X_\mu^a=0 \, ,
\end{equation}
so that the action remains finite. Although this component does \emph{not} affect the global topology, it can become significant locally: we shall see in~\sec{sec:R4-SU2-to-U1} that it can cancel a non-zero $\ell$ twist contribution to the Abelian subcomponent of $A_\mu^a$ on the loop. 
We set $X_\mu^a = 0$ \emph{prior} to the numerical relaxation, though the relaxation algorithm itself will populate this term. Added together, the three components in eq.~\eqref{eq:nonab-A} let $A_\mu^a$ span its entire non-singular field space in the instanton number = 1 topological sector, modulo small gauge transformations that would rotate $\hat{\Phi}$.

Having established our ansatz, we now calculate its instanton number $I$. Since our gauge field ansatz $A_\mu^a$ is free of singularities in the bulk and carries winding number $\ell$ at infinity ($A_\mu^a$ defines a map $S^3_\infty\rightarrow SU(2)=S^3$ that cannot be continuously deformed into the identity map), the instanton number can simply be computed from the Chern-Simons current
\begin{subequations}
\begin{align}
    &K_\mu 
    \equiv\epsilon^{\mu\nu\rho\sigma}\left(A_\nu^a\partial_\rho A_\sigma^a+\frac{1}{3}\epsilon^{abc}A_\nu^a A_\rho^b A_\sigma^c\right)  \label{eq:CSinf1} \\
    &\xrightarrow{u^2 + v^2 \rightarrow \infty} \frac{1}{\left(u^2 + v^2\right)^{3/2}} 
    \begin{pmatrix} 
        \scalebox{0.85}{$2 \cos \vartheta  [(\ell+1) \cos ((1-\ell)\tau -\varphi )+3 (\ell+1) \cos ((1-\ell) \tau +\varphi )-4 \cos \varphi ]$} \\
        \scalebox{0.85}{$2 \cos \vartheta  [-((\ell+1) \sin ((1-\ell) \tau -\varphi ))+3 (\ell+1) \sin ((1-\ell) \tau  +\varphi )-4 \sin \varphi ]$} \\
        \scalebox{0.85}{$8 \ell \sin \vartheta  \cos \tau$}  \\
        \scalebox{0.85}{$8 \ell \sin \vartheta  \sin \tau$}  
    \end{pmatrix} \label{eq:CSinf2} \, .
\end{align}
\end{subequations}
We have written the limit of eq.~\eqref{eq:CSinf2} in terms of $\vartheta = \arctan(u/v) \in [0,\pi/2]$ in anticipation of the next step. 
Using Stokes' theorem to write the instanton number as a surface integral, and integrating over the infinite three-sphere $S^3_{\infty}$ yields
\begin{subequations}
\begin{align}
    I & =\frac{1}{8\pi^2}\int_{\mathbb{R}^4}\mathrm{tr}\left[ G\wedge G\right]  \, , \label{eq:Iasymptotic1}\\
    & = \frac{1}{16\pi^2}\int_{S^3_\infty} \dd S \, \hat{n}_\mu \, K_\mu \, \label{eq:Iasymptotic2} \, ,\\
    &= \frac{1}{16\pi^2}\int_0^{2\pi}\dd\varphi\int_0^{2\pi}\dd\tau\int_0^{\pi/2}\dd \vartheta  \, \cos\vartheta \sin\vartheta  \, \lim_{u^2 + v^2 \to \infty} \left[ \left(u^2+v^2\right)^{3/2} \hat{n}_\mu K_\mu \right] \label{eq:Iasymptotic3}\, ,\\
    &= \ell = \pm 1 \, . \label{eq:Iasymptotic4}
\end{align}
\end{subequations}
In the above, $\hat{n}_\mu = \left(\cos\vartheta  \cos\varphi, \cos \vartheta  \sin \varphi, \sin\vartheta  \cos\tau, \sin\vartheta  \sin \tau \right)$ is the unit normal vector to the $S^3_\infty$ (pointing away from the origin); the surface element $\dd S$ is written in eq.~\eqref{eq:Iasymptotic3}.
This calculation is manifestly independent of the interpolating behavior of the various profile functions determined via numerical relaxation. Hence, our field relaxation procedure cannot alter the non-trivial winding of $A_\mu^a$ at infinity and thus the instanton number, as the procedure only updates in ``small'' steps (not with large gauge transformations).

While our construction generalizes to arbitrary instanton number $I = \ell$, we restrict our numerical analysis in this work to $\ell=\pm 1$ for simplicity. A detailed numerical study of field configurations with higher $\ell$ is left to future work.

\begin{figure}
    \centering
    \includegraphics[width=1.0\linewidth]{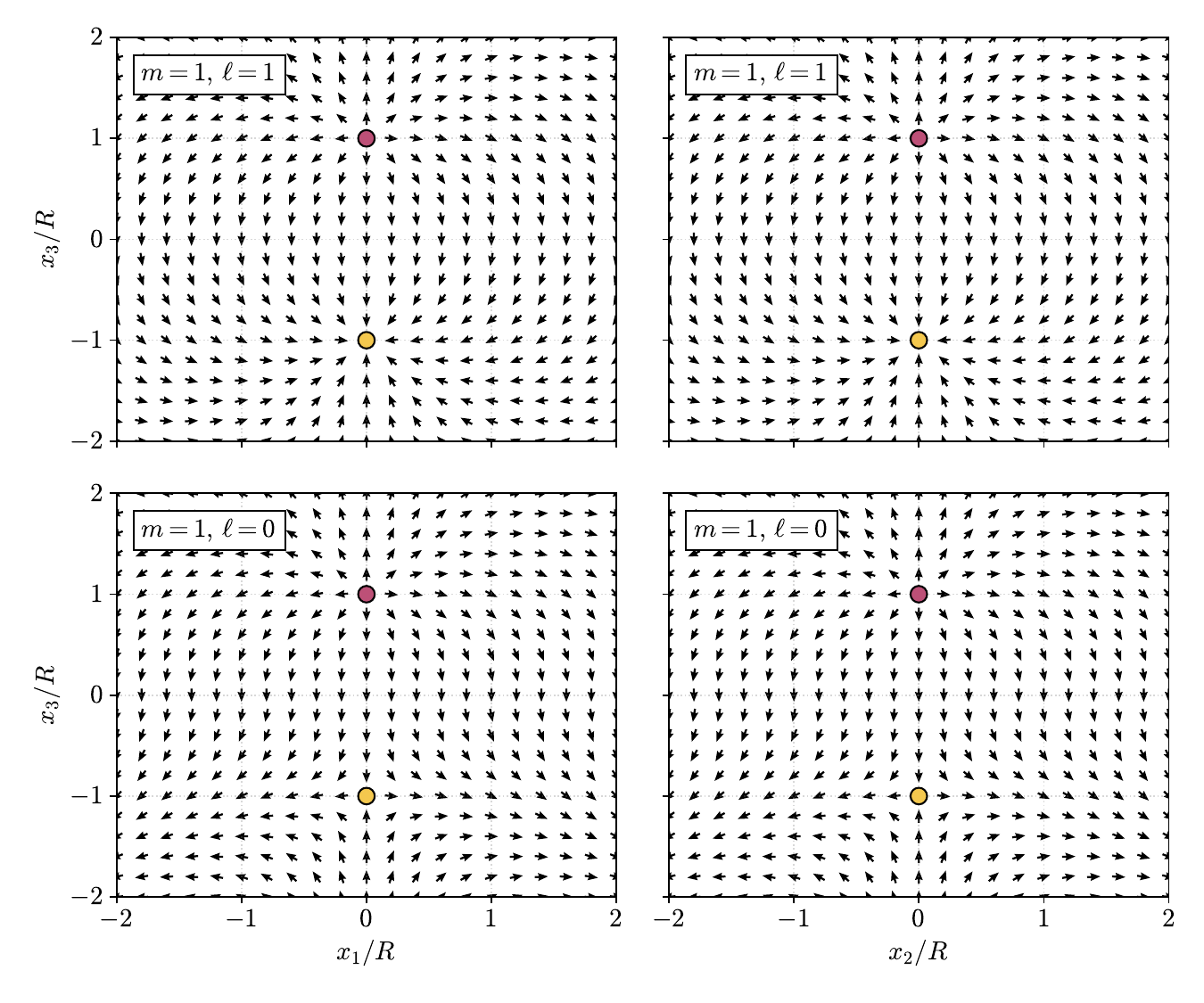}
    \caption{\textit{Top left and right}: The $\ell=0$ Higgs isospin configuration $\hat{\Phi}^a$, given by eq.~\eqref{eq:higgs_isospin}, superimposed onto the Euclidean $x_1x_3$ and $x_2 x_3$ plane, respectively. \textit{Bottom left and right}: The $\ell=1$ Higgs isospin configuration in the Euclidean $x_1 x_3$ and $x_2 x_3$ plane, respectively. Regardless of $\ell$, in this fixed time cross section of the loop, the configuration can be interpreted as a hedgehog-antihedgehog, which corresponds to a monopole-antimonopole pair (or more precisely, a dyon-antidyon pair for $\ell\neq 0$). Note that the third isospin component gets parity flipped with respect to the closest dyon when moving from the upper half $x_3$ (or $x_4$) plane to the lower half $x_3$ (or $x_4$) plane when the isospin is imposed on the spacetime. This is due to a parity flip implicit in the definition of $\beta$. By contrast, there is no implicit parity flip in the first or second components unless $\ell$ is odd. An odd number of isospin components undergoing a parity flip when moving from the upper to the lower half plane means that the field configuration can be interpreted as a dyon-antidyon configuration for any $\ell\in\mathbb{Z}$.}
\label{fig:higgs-isospin-example}
\end{figure}

\subsubsection{Numerical Relaxation}\label{sec:relaxation}

We now outline our numerical relaxation procedure to minimize the action of the specified field configuration. The relaxation code is publicly available in the companion GitHub repository~\githubmaster, with a more detailed description of what follows provided in appendix~\ref{app:numerical_relaxation}. We initialize our ansatz field configuration from eqs.~\eqref{eq:ymh_gauge_field_topology}--\eqref{eq:X_behavior} on a four-dimensional grid, using a fixed loop radius $R$. Without loss of generality, we also set the gauge coupling $g=1$ ($g\neq 0$ configurations are obtained via rescaling). The computational domain extends from $R \pm 1.55\, \Phi_0^{-1}$ in each direction, with a grid spacing of $0.1\, \Phi_0^{-1}$. We iteratively solve the Euclidean Georgi-Glashow equations of motion while holding the gauge and Higgs field values fixed at the boundary (where they asymptote to pure gauge) and keeping the Higgs isospin direction consistent with  eq.~\eqref{eq:higgs_isospin}. 
These conditions amount to restricting the field configuration to the unit-instanton topological sector with a constrained circular loop of a monopole defect in the bulk, at $\lbrace u=0\,, v=R \rbrace $ (which cannot be removed by a non-singular gauge transformation).\footnote{The spacetime grid is constructed so as to avoid the isospin singularities, i.e.~no point lies \emph{exactly} on the loop at $u^2+(v-R)^2 = 0$.} The rigid choice of Higgs isospin direction removes the $SU(2)/U(1)$ part of the gauge redundancy. Apart from these constraints, the fields remain free to evolve.
\begin{figure}
    \centering
    \includegraphics[width=1.0\linewidth]{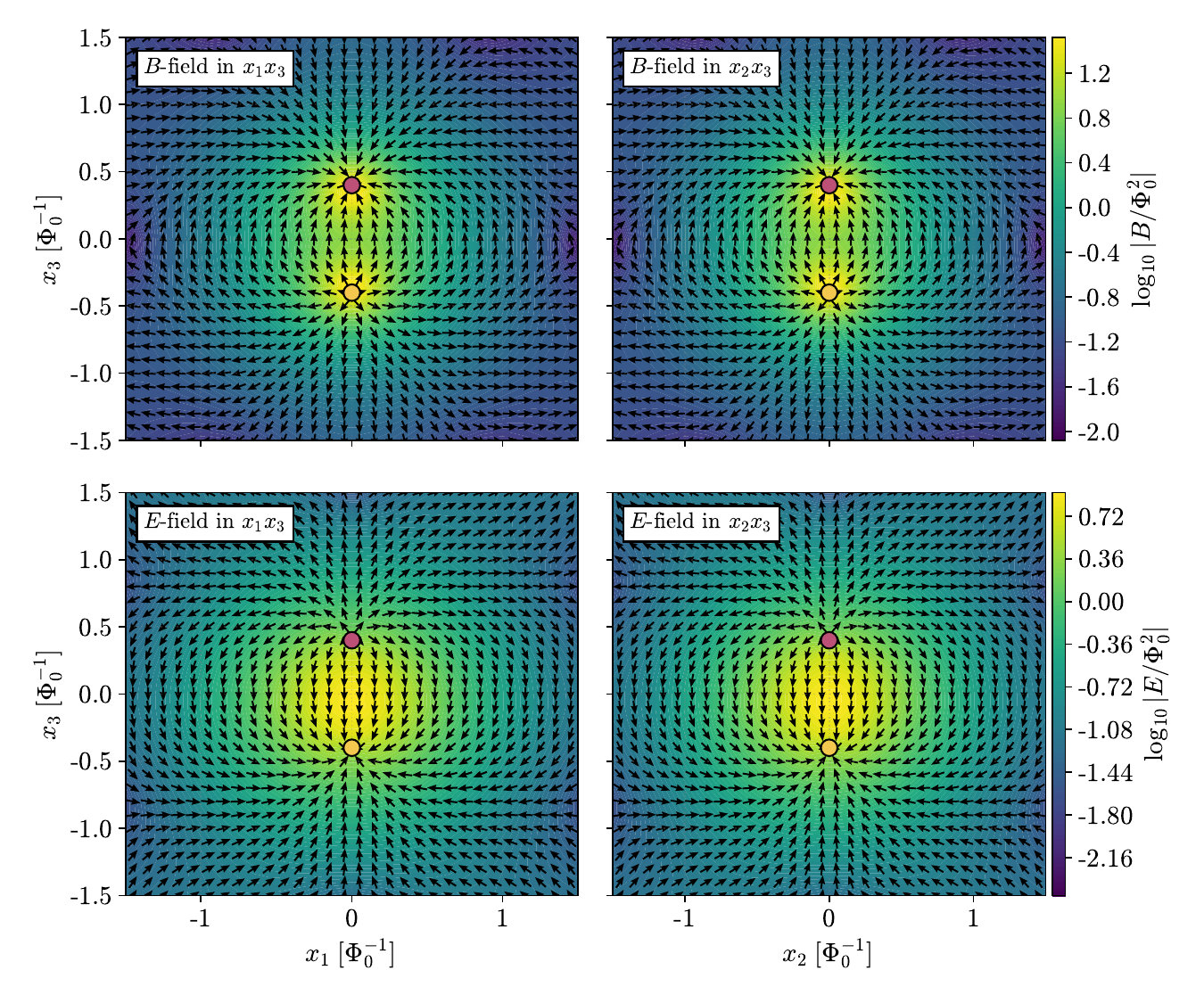}
    \caption{The numerically relaxed $U(1)$ electric and magnetic field for $R=0.4\; \Phi_0^{-1}$ with $\ell=1$, $\lambda=1$, $g=1$. \textit{Top left and top right}: The $U(1)$ magnetic field in the $x_1x_3$ and $x_2x_3$ plane, respectively. \textit{Bottom left and bottom right}: The $U(1)$ electric field in the $x_1x_3$ and $x_2x_3$ plane, respectively. The purple and yellow dots show the Higgs isospin loci, defined by eq.~\eqref{eq:higgs_isospin}. The arrows show the local field direction, which unlike the Higgs isospin direction is not held fixed, while the color indicates the logarithm of the field magnitude. These $E$ and $B$ fields, which are embedded in a larger $SU(2)$ field strength, are defined by eq.~\eqref{eq:thooft_tensor}. At this loop radius, the electric and magnetic fields resemble that of a dyon-antidyon pair, consisting of point-like magnetic charges and a smooth electric charge profile.}
\label{fig:relaxed_dyon_abelian_strength_small_R}
\end{figure}
\begin{figure}
    \centering
    \includegraphics[width=1.0\linewidth]{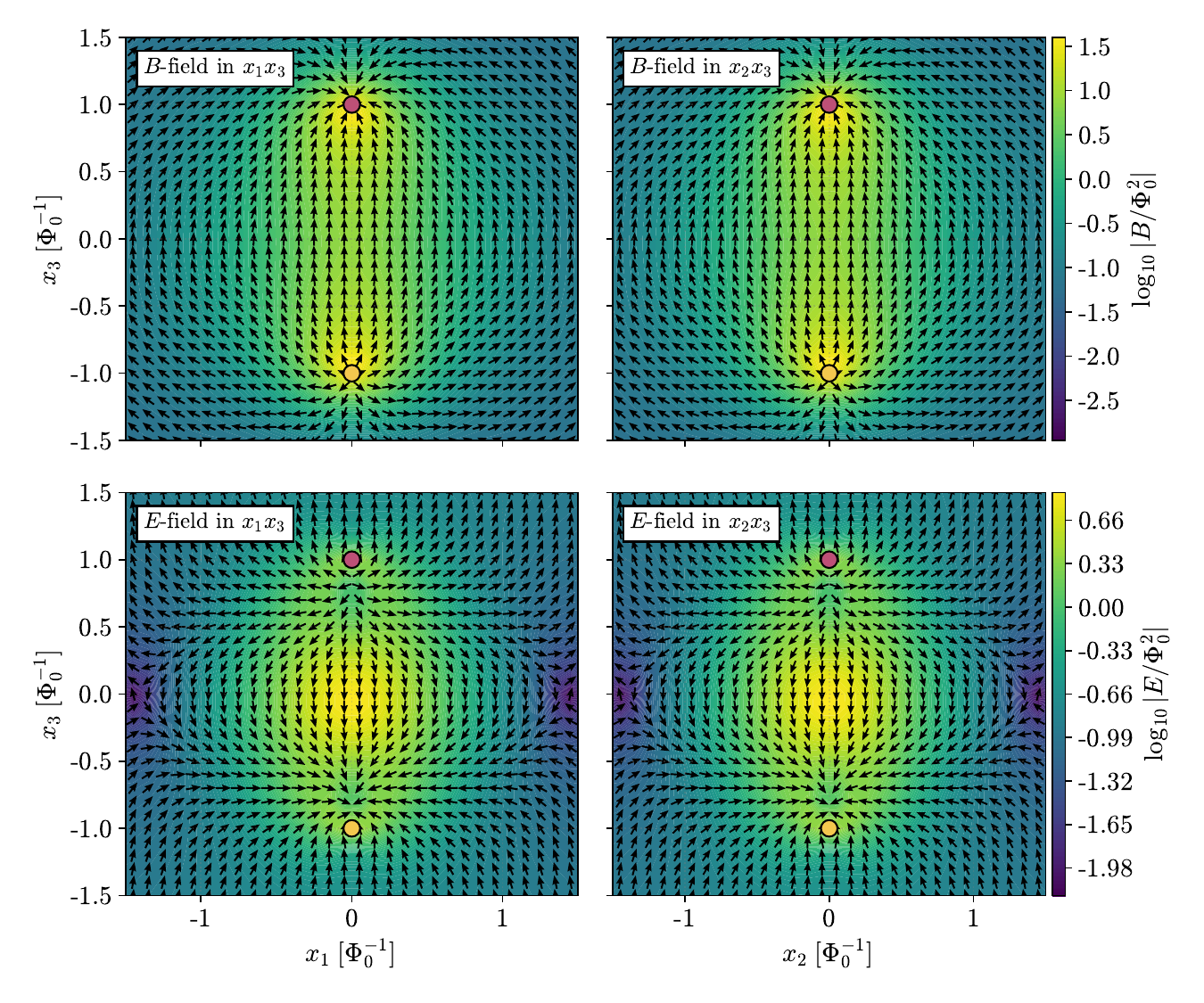}
    \caption{The numerically relaxed $U(1)$ electric and magnetic field for $R=1.0\; \Phi_0^{-1}$ with $\ell=1$, $\lambda=1$, $g=1$. \textit{Top left and top right}: The $U(1)$ magnetic field in the $x_1x_3$ and $x_2x_3$ plane, respectively. \textit{Bottom left and bottom right}: The $U(1)$ electric field in the $x_1x_3$ and $x_2x_3$ plane, respectively. The purple and yellow dots show the Higgs isospin loci, defined by eq.~\eqref{eq:higgs_isospin}. The arrows show the local field direction, which unlike the Higgs isospin direction is not held fixed, and are unit length, while the color indicates the logarithm of the field magnitude. These $E$ and $B$ fields, which are embedded in a larger $SU(2)$ field strength, are defined by eq.~\eqref{eq:thooft_tensor}. At this larger loop radius, the magnetic field still emerge from point sources but are more columnated than one would expect from a pure $U(1)$ monopole. The electric fields also resemble that a $U(1)$ dipole configuration, but with the arrows converging closer to the center of the loop, hence corresponding to an electric charge distribution that is peaked away from the monopole point source. This shows that fixing Higgs loci at $u=0$, $v=R$ is not enough to make the extremal action field configuration become fully dyon-loop like --- with the more centered field strength components being a characteristic of the standard instanton.}
\label{fig:relaxed_dyon_abelian_strength_large_R}
\end{figure}
We define convergence as the point when no field value changes by more than $10^{-5} \, \Phi_0$ per iteration on the grid. The relaxation algorithm is run for loop radii $R\in\{0.0,0.1,\dots,1.0\} \, \Phi_0^{-1}$ and coupling constants $\lambda\in\{0.0,0.5,1.0\}$. 
Upon convergence, we obtain field configurations that minimize the Euclidean Georgi-Glashow action. 

\begin{figure}
    \centering
    \includegraphics[width=1.0\linewidth]{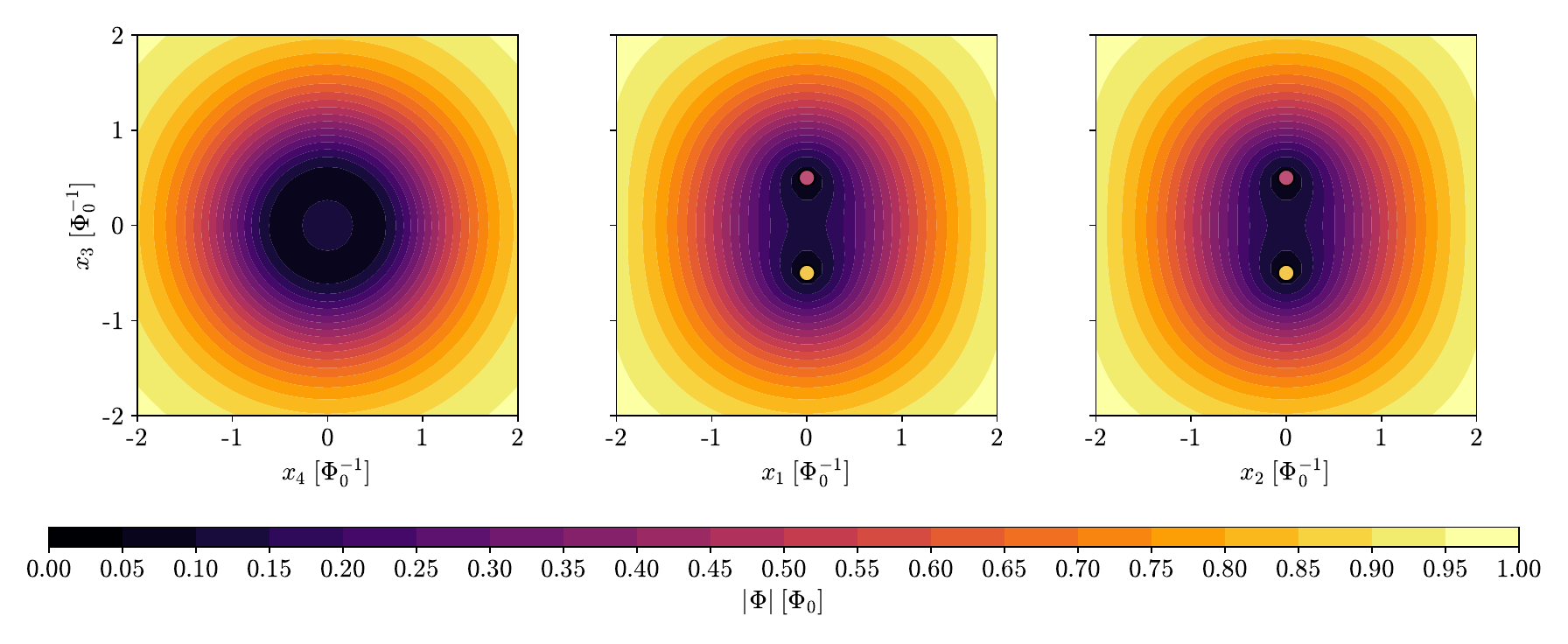}
    \caption{Three numerically relaxed Higgs field cross sections for $R=0.5\, \Phi_0^{-1}$ with $\lambda=1$, $g=1$. \textit{Left}: The $x_3 x_4$ plane, coinciding with the loop. \textit{Center and right}: Cross sections taken through the loop in the $x_1 x_3$ and $x_2 x_3$ plane, respectively. The purple and yellow dots show the Higgs isospin loci, defined by eq.~\eqref{eq:higgs_isospin}. In every slicing, the Higgs field drops to zero in the core of the loop. Outside the core, the Higgs field rapidly asymptotes to its vacuum expectation value $|\Phi|\rightarrow\Phi_0$, while in the center, it takes on an intermediate value. The location of the Higgs zero points are dictated by the fixed $\hat{\Phi}$, which fixes the physical hedgehog loop defect while being a pure gauge choice away from the locus $u^2 + (v-R)^2 = 0$.}
\label{fig:higgs_norm}
\end{figure}

\begin{figure}
    \centering
    \includegraphics[width=0.9\linewidth]{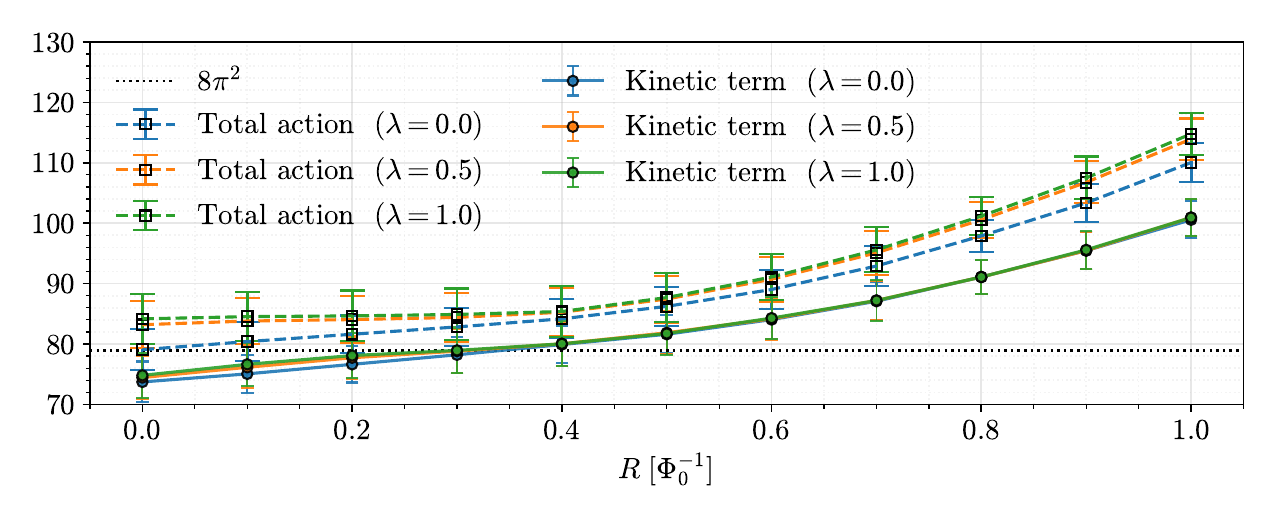}
    \caption{The numerically relaxed Georgi-Glashow dyon loop action across various loop radii, assuming $g=1$,  $\ell=1$. The dashed curves show the total Euclidean action $S$, while the solid curves show the integrated kinetic term $\frac{1}{2g^2}\int \mathrm{tr}[G\wedge\star G]$. Blue, orange, and green markers correspond to $\lambda=0$, $\lambda=0.5$, and $\lambda=1$, respectively. The error bars reflect numerical uncertainty --- they are computed from i) the error due to the integrated action density \textit{outside} the $2(R+1.55\Phi_0^{-1})$ box size and ii) the error due to finite lattice spacing, computed using a second-order Richardson error estimate. The black dotted line shows $8\pi^2$ to enable comparison to the BPST instanton action. The total action and gauge kinetic action are both monotonically increasing as a function $R$, with the increase in action being especially prevalent for $R\sim \Phi_0^{-1}$. At $R= 0$, the gauge kinetic action is approximately $4 \%$ below the BPST instanton gauge field action for all values of $\lambda$, largely consistent with the numerical error estimate. In agreement with our analytic proof in appendix~\ref{app:no-other-action-minimum}, we find no evidence for a local action minimum at $R>0$, meaning the only stable configuration in this topological sector is the Higgsed BPST instanton. However, we note that the relatively low action slope at $R\ll \Phi_0^{-1}$ means that a small dyon loop is a good approximation of the instanton configuration, and that such configurations may easily be realized by quantum perturbations.}
\label{fig:action}
\end{figure}

Although our initial guess differs significantly from a pure dyon loop, the relaxed configuration clearly exhibits a dyon dipole structure. This is illustrated in figure~\ref{fig:relaxed_dyon_abelian_strength_small_R}, where we plot the Abelian $E$ and $B$ fields associated with the 't Hooft tensor of eq.~\eqref{eq:thooft_tensor}. 
Magnetic sources clearly appear at $z=\pm R$, sourced by the term $-\epsilon^{abc}D_\mu\hat{\Phi}^bD_\nu\hat{\Phi}^c$. 
The electric charge distribution is \emph{not} topologically constrained --- it is determined by dynamics, i.e.~the equations of motion --- and peaks slightly closer to the center of the loop, especially for large loop radii, as shown in figure~\ref{fig:relaxed_dyon_abelian_strength_large_R}.
Figure~\ref{fig:higgs_norm} shows the norm of the Higgs triplet, demonstrating that it vanishes at the monopole core, restoring the unbroken $SU(2)$ phase locally, and smoothly approaches the vacuum expectation value $\Phi_0$ towards spatial infinity.
In figure~\ref{fig:action}, we plot the gauge field action and total action as a function of loop radius $R$. 
The action's global minimum occurs at $R=0$;  no additional local minima are observed. This agrees with our analytic proof that the Georgi-Glashow model does not support any smooth semiclassical extremum of the action other than the zero-size instanton in appendix~\ref{app:no-other-action-minimum}. In the zero-radius limit, the gauge field action matches the BPST instanton action $S_\mathrm{gauge}=8\pi^2/g^2$ within numerical precision, confirming that the dyon loop configuration continuously reduces to the standard Higgsed BPST instanton as $R\rightarrow0$. Although the true minimal action configuration for the Georgi-Glashow model would prefer an infinitely small instanton,\footnote{The finite resolution of our numerical implementation artificially limits the smallest size of the instanton to be of order the grid spacing.} the negligible action difference for finite but small $R$ suggests that dyon loops with $R\lesssim \Phi_0^{-1}$ closely approximate the minimal-action configuration. Crucially, while a regular instanton has a point isospin singularity, our dyon loop configuration features a loop singularity, a topological distinction that becomes essential in singular gauge, where Abelian dyon loop configurations emerge naturally only if supported by such loop singularities.

\subsection{Abelian Reduction}\label{sec:R4-SU2-to-U1}
We now take our dyon loop solution to singular gauge, allowing clear separation between the Abelian and other gauge field components, parameterized by $\mathcal{A}$ and $W$, respectively, as done previously in eq.~\eqref{eq:singular_separation}. 
This facilitates the direct analysis of the Abelian contribution to the instanton number, analogous to our earlier discussion of the free dyon line in $\mathbb{R}^3 \times S^1$ of \sec{sec:free_dyon_su2u1}.

To achieve singular gauge, we apply the gauge transformation
\begin{align}
    \Omega = e^{i \gamma \sigma_3/2} e^{i \beta \sigma_2 / 2} e^{i \alpha \sigma_3/2} \, , 
\label{eq:singular-gauge-transf-R4}
\end{align}
with $\alpha = \varphi + \tau$ and $\beta$ defined in eq.~\eqref{eq:beta}. The angle $\gamma$ parameterizes a rotation around the residual unbroken $U(1)$ and determines the location of the Dirac sheet; choosing $\gamma = -\varphi+\tau$ places the sheet on the disk $D_\leq$ shown in figure~\ref{fig:sheets}. This transformation unwinds the non-trivial Hopf map at infinity and aligns the Higgs field with the third isospin direction everywhere, i.e.~$\hat{\Phi}^a(x) = \delta^{a3} \, \forall \, x \in \mathbb{R}^4$. Since the Higgs and gauge field are ``matched'' at infinity in the sense that $\lim_{u^2+v^2\rightarrow\infty}(D_\mu\Phi)^a=0$ (necessary to keep the action bounded), this also removes the non-trivial gauge field instanton winding on $S^3_\infty$. Consequently, the winding at infinity is transferred to a winding around the singular loop at the locus $u^2 + (v-R)^2 = 0$.

In singular gauge, the Abelian gauge field is given by
\begin{subequations}
\label{eq:pure-gauge-piece}
\begin{align}
\mathcal{A}&=\frac{1}{2}a +\frac{1}{2}\left[\cos\beta\;\dd\alpha+\dd\gamma\right]+\frac{1}{2}X^a\hat{\Phi}^a\, \\[5pt]
&=\frac{1}{2}a +\frac{1}{2}\left[\left(\cos\beta-1\right)\dd\varphi+(\cos\beta+1)\dd\tau\right] + \frac{1}{2}X^a\hat{\Phi}^a\, ,
\end{align}
\end{subequations}
where $a$ is the Higgs-aligned gauge field component of $A^a$, where $X^a$ is the topologically trivial gauge field component of $A^a$ (both defined in~\sec{sec:ansatz}) and where $\hat{\Phi}^a$ is the \emph{unrotated} (regular-gauge) Higgs isospin map defined by eq.~\eqref{eq:higgs_isospin}.
Similarly, the $W$-boson field is
\begin{subequations}
\label{eq:W_pure-gauge-piece}
\begin{align}
W&=\frac{1}{\sqrt{2}}\left[1-K(u,v)\right]e^{i(\varphi-\tau)}\left(i\dd\beta-\sin\beta\dd\varphi-\sin\beta\dd\tau\right)+\frac{1}{\sqrt{2}}\left(\tilde{X}^1+i\tilde{X}^2\right)\, ,
\end{align}
\end{subequations}
where $\tilde{X}=\Omega X \Omega^\dagger$ is the gauge rotated $X$ component from~\sec{sec:ansatz}. Comparing eq.~\eqref{eq:pure-gauge-piece} and eq.~\eqref{eq:W_pure-gauge-piece} with the Abelian gauge field configuration in~\sec{sec:abelian-loop}, we see identical singular structures, although with additional $\cos\beta \,\dd\tau$, $X^a\hat{\Phi}^a$, and $\tilde{X}^1+i\tilde{X}^2$ terms. The $\mathcal{A}_\varphi$ component notably also has a flipped sign, indicating we have a monopole with negative magnetic charge. For small loop radii ($R\lesssim \Phi_0^{-1}$), numerical relaxation shows that $X^a \approx -\delta^{a3} \dd\tau$, exactly canceling the extra $\cos\beta$ term and matching the construction in \sec{sec:abelian-loop}.
For larger loops ($R\gtrsim \Phi_0^{-1}$), this cancellation is incomplete, leading to an offset in the electric charge distribution relative to the loop, as depicted in figure~\ref{fig:relaxed_dyon_abelian_strength_large_R}.
Imposing the artificial requirement that $X^a$ take on this specific canceling value on the loop to retain a ``pure dyon'' loop does not follow from topological considerations and is disfavored by the least-action principle within this specific UV completion.

We now evaluate the instanton number in singular gauge using the result obtained in eq.~\eqref{eq:I_string_gauge_tot}. Like in~\sec{sec:free_dyon_su2u1}, we do this by using Stokes' theorem to convert the bulk instanton number integrals to integrals along surfaces where $\mathcal{A}$ and $W$ as specified by eq.~\eqref{eq:pure-gauge-piece} and eq.~\eqref{eq:W_pure-gauge-piece}, respectively, become singular. The $W$ field is only singular on the dyon worldline. Since $X$ is smooth and can take on any value on the worldline, we choose to set it to zero in the subsequent topological calculations without loss of generality. In this case, the $W$ contribution to the instanton number vanishes trivially since
\begin{equation}
    (W\wedge D W^*)_{\beta\varphi\tau}\Big|_\mathrm{loop} = 0 \, ,     
\end{equation}
analogously to~\sec{sec:free_dyon_su2u1}. In this singular gauge, the instanton number thus resides entirely in the Abelian sector. The Abelian instanton number can be computed via two surface integrals: one on the $S^2_\varepsilon\times S^1$ wrapping the monopole worldline, and another on the $[0,R)\times S^1\times S^1$ surface wrapping the Dirac sheet (see figure~\ref{fig:summary} for a visual depiction of the Dirac sheet and the surface wrapping the loop). Specifically, when $X^3=0$, the instanton number can be computed entirely from the worldline surface
\begin{subequations}
\begin{align}
    -\frac{1}{4\pi^2}\int_{S^1\times S^2_\varepsilon} \mathcal{A}\wedge\mathcal{F}&=-\frac{1}{4\pi^2}\int_0^{\pi}\dd\beta\int_0^{2\pi}\dd\varphi\int_0^{2\pi}\dd\tau \, \left(\mathcal{A}\wedge\mathcal{F}\right)_{\beta\varphi\tau}\\[5pt]
    &=\frac{1}{4\pi^2}\int_0^{\pi}\dd\beta\int_0^{2\pi}\dd\varphi\int_0^{2\pi}\dd\tau \, \left(\frac{1}{2}\sin\beta\right)
    = +1 \, ,
\end{align}
\end{subequations}
where the overall minus sign on the left hand side comes from the orientation induced by the $\dd^4 x$ volume element. For other values of $X^3$ on the loop, contributions shift between the loop and the Dirac sheet, but the total instanton number remains unchanged. Altogether, we conclude that the instanton number can be computed entirely from the Abelian gauge field as
\begin{equation}
    I = \frac{1}{4\pi^2}\int_{\mathbb{R}^4} 
    \mathcal{F}\wedge\mathcal{F}=\frac{1}{4\pi^2}\int_{\Sigma(\mathrm{singular\;surfaces})} \mathcal{A}\wedge\mathcal{F}=+1 \, ,
\end{equation}
with the plus sign coming from the dyon having negative magnetic charge and positive electric charge. This result clearly connects to \sec{sec:abelian-loop}: the $U(1)$ instanton configuration emerges as the Abelian subcomponent of an $SU(2)$ gauge field, fully separated from the non-Abelian degrees of freedom. The price of this separation is the appearance of singularities on the loop and the Dirac sheet. These defects are precisely what enable the Abelian gauge field $\mathcal{A}$ to carry the instanton number, supporting the interpretation of the configuration as an Abelian instanton in the infrared.
This insight is particularly intriguing, as we showed in \sec{sec:non-abelian-loop} that the dyon loop is a continuous deformation of the standard instanton. Therefore, this opens up the possibility of approximate calculations of non-perturbative effects without detailed knowledge of the UV completion, for instance relying solely on properties such as the dyon mass rather than its precise internal structure and interactions.

\section{Conclusions and Outlook} \label{sec:conclusions}

We have constructed bottom-up field configurations of Abelian gauge fields that carry non-zero instanton number. These Abelian instantons are supported by closed monopole worldlines embedded in $\mathbb{R}^4$ and require the $U(1)$ gauge field to wind non-trivially around the loop. Introducing such a monopole loop modifies the space on which the $U(1)$ gauge theory is defined, turning it into $\mathbb{R}^4 \setminus S^1$. The latter is a topologically non-trivial space with an internal boundary $S^1 \times S^2_\varepsilon$ surrounding the monopole loop. Simultaneous winding of $A$ along the $S^1$ and wrapping of $F$ around the $S^2_\varepsilon$ are the necessary ingredients to obtain non-zero instanton number, proportional to $\int F \wedge F$. These configurations describe Euclidean dyon loops, characterized by localized magnetic charge and smoothly distributed electric charge, explicitly realizing the physical scenario discussed in ref.~\cite{Fan:2021ntg}.

We have further demonstrated that these Abelian instantons admit a UV completion within the Georgi-Glashow model as loops of Julia-Zee dyons. 
Remarkably, we found that the instanton number in these UV-complete solutions is carried entirely by the unbroken Abelian component, a feature we showed even holds for the ``free" Julia-Zee dyon solutions, whose (non-Abelian) instanton number was previously studied in  refs.~\cite{Marciano:1976as,Christ:1979iw}. This crucial property allows these field configurations to be identified in the Abelian effective theory, despite the finer structural details requiring knowledge of the UV completion. Implementing a numerical relaxation algorithm, we confirmed that a Euclidean Julia-Zee dyon loop is a continuous deformation of the small constrained instanton present in the Georgi-Glashow model~\cite{Affleck:1980mp}.

Our findings suggest several avenues for future work. 
First, the calculation of the axion potential described in ref.~\cite{Fan:2021ntg} merits further refinement. Our findings are consistent with using a semiclassical expansion around zero dyon loop radius (with the gauge field winding being non-trivial) as a way to approximate standard instanton effects. However, an exact mapping between effects calculated from standard instanton methods and effects calculated by means of a dyon loop expansion should be explored further. Furthermore, the purely Abelian nature of the instanton number associated with the dyon loops suggests the possibility of systematically parameterizing axion potential contributions (and potentially other physical effects) in terms of monopole masses and dyon excitation energies. Such an approach could significantly enhance the predictive power of calculations involving instantons.

Second, exploring the influence of light fermions on these Abelian instantons is critical for many phenomenological applications. Even in the presence of magnetic monopoles, an Abelian $\theta$ term will be unphysical in the presence of massless fermions and, as a result, Abelian instanton contributions to physical observables must be suppressed by a power of the light fermion mass. Understanding how this fermion mass dependence plays out in the field configurations described here is a natural next step. A third intriguing direction is to generalize the single-loop construction studied here. While our analysis in \sec{sec:non-abelian-loop} focused on unit-twist ($|\ell|=1$) dyon configurations, relating directly to single instantons, the structure of solutions with $|\ell|>1$, presumably related to multi-instanton configurations, remains an open question that could reveal new and distinct behavior.

Finally, examining alternative UV completions of Abelian instantons presents an exciting possibility. 
The 't Hooft-Polyakov monopole is the simplest field-theoretic resolution of the Dirac monopole singularity, but alternative possibilities exist in the context of larger non-Abelian gauge groups, theories with extra-dimensions, or even string theory. An analysis of these Abelian instantons in the context of four-dimensional Einstein-Maxwell theory will appear in forthcoming work~\cite{EMGR_upcoming}. More generally, for consistent coupling to e.g. the axion, the UV monopoles must satisfy certain requirements which have previously been studied in the context of anomaly inflow~\cite{Fukuda:2020imw} and the gauging of Chern-Weil symmetries~\cite{Heidenreich:2020pkc,Garcia-Valdecasas:2024cqn}. It would be interesting to understand these consistency conditions in terms of the relation between standard UV instantons and the Euclidean dyon loop constructed in this work.

In summary, the field theory and potential phenomenological applications of Abelian instantons are very rich. Many intriguing questions remain open, and we hope this study sparks further exploration into the properties and implications of these non-perturbative objects.

\acknowledgments
We thank Prateek Agrawal, Aleksey Cherman, Nathaniel Craig, Patrick Draper, Fabian Hahner, Anson Hook, Elliot Maderazo, John March-Russell, Jacob McNamara, Fedor Popov, Arkady Vainshtein, Lian-Tao Wang, and Yifan Wang for useful conversations, and David Dunsky, Anson Hook, Junwu Huang, Matt Reece, and Grant Remmen for useful comments on the manuscript.
The research of IGG is supported by the U.S.~Department of Energy grant No.~DE-SC0011637.
MK~is supported by a James Arthur Graduate Assistantship at New York
University.
The research of KVT is supported by the National Science Foundation under Grant No.~PHY-2210551. The Center for Computational Astrophysics at the Flatiron Institute is supported by the Simons Foundation. 
This work was supported in part through the NYU IT High Performance Computing resources, services, and staff expertise. The numerical section of this work used the Python packages \texttt{numpy}~\cite{harris2020array}, \texttt{scipy}~\cite{2020SciPy-NMeth}, and \texttt{JAX}~\cite{jax2018github}.
\appendix

\section{Abelian Instanton Action}\label{sec:app_action}

In this appendix, we fill in the details behind the action calculation of section~\ref{sec:u1action}. Although the instanton field configuration described in section~\ref{sec:abelian-loop} is qualitatively different from the bounce describing the decay of a homogeneous magnetic field into monopole pairs that is the topic of ref.~\cite{Affleck:1981ag}, the calculation of the corresponding Euclidean action proceeds very much along the same lines. Below, we closely follow ref.~\cite{Affleck:1981ag} in estimating the instanton action.

Let us first discuss the second term in eq.~\eqref{eq:U1action_split}. To regulate this quantity, we introduce a smooth magnetic charge density, analogous to the electric charge density of eq.~\eqref{eq:jeloop}, so that eq.~\eqref{eq:jmonopoleloopv1} is now replaced by
\begin{align}
    j_m \rightarrow \rho_m (u, v) \left( - \sin \tau \dd x_3 + \cos \tau \dd x_4 \right) \ ,
\end{align}
with the restriction 
\begin{equation} \label{eq:mR2}
    m R^2 = - \iint \dd u \dd v \, uv^2 \rho_m(u,v) \ .
\end{equation}
In the following, we assume that both the electric and magnetic charge densities are localized within some distance $l$ of the loop at $u=0$, $v=R$, and that this loop ``thickness" is $l \sim l_e\sim l_m \ll R$. Working in Lorentz gauge, $\Delta S_\text{IR}$ can then be written as follows
\begin{align}
    \Delta S_\text{IR} & = \frac{1}{4 \pi^2 e^2} \int \dd^4x \, \dd^4x' \left\{ \frac{j_{m, \mu} (x) j_{m, \mu} (x')}{(x - x')^2} + [ j_m \rightarrow e^2 j_e ] \right\} \\
    & = \frac{1}{4 \pi^2 e^2} \int \dd^4x \, \dd^4x' \, \frac{\cos (\tau - \tau')}{(x - x')^2} \left\{ \rho_m (u,v) \rho_m (u',v') + [ \rho_m \rightarrow \rho_e ] \right\} \ .
\end{align}
Performing the integrals over $\tau$, $\tau'$, the above expression then reads
\begin{align}
    \Delta S_\text{IR} = & - \frac{1}{2e^2} \left( \int \dd x_1 \dd x_2 \dd v \rho_m (u,v) \right)^2  + \left[ \rho_m \rightarrow \rho_e \right] \label{eq:SIR_charge2}\\
    & + \frac{1}{2e^2} \int \dd x_1 \dd x_2 \dd v \int \dd x'_1 \dd x'_2 \dd v' \rho_m (u,v) \rho_m (u',v') \frac{W}{\sqrt{W^2 - 1}} + \left[ \rho_m \rightarrow \rho_e \right] \ , \label{eq:SIR_chargecharge} \\
    & \text{with} \qquad W \equiv \frac{1}{2 v v'} \left( (x_1 - x'_1)^2 + (x_2 - x'_2)^2 + v^2 + v'^2 \right) \ .
\end{align}
Under the assumption that the magnetic and electric charge densities are highly localized near the dyon loop, the quantity inside parenthesis in eq.~\eqref{eq:SIR_charge2} is approximately given by
\begin{equation}
    \int \dd x_1 \dd x_2 \dd v \rho_m (u,v) \simeq \frac{2 \pi}{R^2} \int \dd u \dd v \, u v^2 \rho_m (u,v) = - 2 \pi m \ ,
\end{equation}
by virtue of eq.~\eqref{eq:mR2}, and similarly for the integral involving $\rho_e$. Eq.~\eqref{eq:SIR_charge2} is then approximately given by
\begin{equation}
    - \frac{1}{2e^2} \left( \int \dd x_1 \dd x_2 \dd v \rho_m (u,v) \right)^2  + \left[ \rho_m \rightarrow \rho_e \right] \simeq - \frac{2 \pi^2 (m^2 + q^2)}{e^2} \ .
\end{equation}
To estimate eq.~\eqref{eq:SIR_chargecharge}, we expand the quantity $W / \sqrt{W^2-1}$ near $x_1 = x_2 =0$ and $v=R$, as follows
\begin{equation} \label{eq:W}
    \frac{W}{\sqrt{W^2-1}} = \frac{R}{|{\bf r} - {\bf r'}|} + \frac{z+z'}{2 |{\bf r} - {\bf r'}|} + \mathcal{O} \left( \frac{r, r'}{R} \right) \ ,
\end{equation}
where $z \equiv v-R$ and ${\bf r} \equiv (x_1, x_2, z)$, and similarly for $z'$ and ${\bf r'}$. Eq.~\eqref{eq:SIR_chargecharge} then reads
\begin{align}
    & \frac{1}{2e^2} \int \dd x_1 \dd x_2 \dd v \int \dd x'_1 \dd x'_2 \dd v' \rho_m (u,v) \rho_m (u',v') \frac{W}{\sqrt{W^2 - 1}} \label{eq:aux1}\\
    & = \frac{1}{2e^2} \int \dd x_1 \dd x_2 \dd z \int \dd x'_1 \dd x'_2 \dd z' \rho_m (u,z) \rho_m (u',z') \left( \frac{R}{|{\bf r} - {\bf r'}|} + \frac{z+z'}{2 |{\bf r} - {\bf r'}|} + \mathcal{O} \left( \frac{r, r'}{R} \right) \right) \label{eq:aux2}\\
    & \simeq 2 \pi R \times \frac{1}{4\pi e^2} \int \dd^3 {\bf r} \int \dd^3 {\bf r'} \frac{\rho_m ({\bf r}) \rho_m ({\bf r'})}{|{\bf r} - {\bf r'}|}
    + \int \dd^3 {\bf r} \int \dd^3 {\bf r'} \rho_m ({\bf r}) \rho_m ({\bf r'}) \mathcal{O} \left( \frac{r, r'}{R} \right) \ . \label{eq:aux3}
\end{align}
The coefficient of $2 \pi R$ in the first term above is precisely the static self-energy of the charge density $\rho_m$, which diverges in the limit of zero monopole thickness. The second term inside the parenthesis in eq.~\eqref{eq:aux2} vanishes upon integration over $z$ and $z'$, up to exponentially small corrections, for the following reason: since $\rho_m$ is exponentially localized near $z=0$, one might extend the limits of integration $[-R, \infty) \rightarrow (-\infty, \infty)$. With the regime of integration now being symmetric, terms that are odd in $z$ and $z'$ integrate to zero, explaining the absence of $\mathcal{O} \left( R^0\right)$ terms in eq.~\eqref{eq:aux3}. Finally, the last term in eq.~\eqref{eq:aux3} is of order $l^2 \Lambda/R$. Assuming that $l \propto \Lambda^{-1}$, $l^2 \Lambda/R \propto 1 / (\Lambda R)$, and so this correction will vanish as $\Lambda \rightarrow \infty$.
In total, we can then write $\Delta S_\text{IR}$ as in eq.~\eqref{eq:DeltaSIR}, with $\Lambda$ referring to the Coulomb self-energy of our regulated dyon configuration, i.e.~
\begin{equation} \label{eq:Lambda_def}
    \Lambda \equiv \frac{1}{4\pi e^2} \int \dd^3 {\bf r} \int \dd^3 {\bf r'} \frac{\rho_m ({\bf r}) \rho_m ({\bf r'})}{|{\bf r} - {\bf r'}|} + [\rho_m \rightarrow \rho_e] \ .
\end{equation}

We now turn to the first term in eq.~\eqref{eq:U1action_split}. To estimate the size of $\Delta S_\text{UV}$, we will similarly make the assumption that the UV-complete description of our dyon loop only differs from the purely Abelian description within a distance $l \ll R$ from the $S^1$. The lengthscale $l$ can be thought of as the radius of the dyon excitation in the UV theory, and the requirement that $l \ll R$ will restrict the regime of validity of our result. Further assuming that the instanton solution in the full UV theory shares the azimuthal symmetry of our Abelian dyon loop, we have
\begin{align}
    \Delta S_\text{UV}
    & = 2 \pi \iint_{-\infty}^{+\infty} \dd x_1 \dd x_2 \int_0^{\infty} \dd v v \, \left\lbrace \mathcal{L}_\text{UV} - \frac{1}{4e^2} F_{\mu \nu} F_{\mu \nu} \right\rbrace \label{eq:SUV_1}\\
    & = 2 \pi R \iint_{-\infty}^{+\infty} \dd x_1 \dd x_2 \int_{-R}^\infty \dd z \left( 1 + \frac{z}{R} \right) \, \left\lbrace \mathcal{L}_\text{UV} - \frac{1}{4e^2} F_{\mu \nu} F_{\mu \nu} \right\rbrace \label{eq:SUV_2}\\
    & \simeq 2 \pi R \iiint_{-\infty}^{+\infty} \dd x_1 \dd x_2 \dd z \left( 1 + \frac{z}{R} \right) \, \left\lbrace \mathcal{L}_\text{UV} - \frac{1}{4e^2} F_{\mu \nu} F_{\mu \nu} \right\rbrace \Big|_{S^1} \ . \label{eq:SUV_3} \\
     & \simeq 2 \pi R \left( M - \Lambda \right) + \mathcal{O}\left( \frac{l^2 M}{R} \right) \ . \label{eq:SUV_4}
\end{align}
In going from eq.~\eqref{eq:SUV_1} to \eqref{eq:SUV_2} we have introduced a new coordinate $z \equiv v-R$, just as below eq.~\eqref{eq:W}. As in the preceding paragraph, the limits of integration over $z$ may be extended to include the entire real line, at the cost of exponentially small errors, and we may expand the integrand around $x_1=x_2=z=0$. Terms that are odd under $x_1$, $x_2$ and $z$ will vanish upon integration. The leading term proportional to $2\pi R$ coming from the expansion of the UV theory is -- by definition -- the dyon mass in the limit of zero thickness, which we denote as $M$. The Abelian kinetic term contributes to eq.~\eqref{eq:SUV_3} just as in eq.~\eqref{eq:aux3} up to the overall sign being now negative.

\section{Numerical Relaxation Scheme}\label{app:numerical_relaxation}
In this appendix, supplementing \sec{sec:non-abelian-loop}, we
describe the numerical scheme used to obtain field configurations that extremize the Georgi-Glashow action~\eqref{eq:georgi_glashow_action} subject to the constraint that the field configuration contains a monopole loop defect at a fixed radius $R$. The code used to carry out the numerical relaxation scheme is found on the GitHub repository~\githubmaster accompanying this work. We utilize the Jacobi method to solve the Euclidean field equations on a lattice. The analytic equations of motion are
\begin{align}
    \left(D_\mu D_\mu \Phi\right)^a &= \lambda \left(\Phi^b \Phi^b - \Phi_0^2\right) \Phi^a \, , \\
    D_\mu G^a_{\mu \nu} &= g \epsilon^{abc} \Phi^b \left(D_\mu \Phi\right)^c \label{eq:EOM_phi} \, .
\end{align}
To solve these equations numerically, we need to expand them in explicit detail. The expanded Higgs equations of motion are
\begin{multline}
     \lambda\hat{\Phi}^a|\Phi|^3 =(\partial_\mu\partial_\mu\hat{\Phi}^a+g^2A_\mu^a A_\mu^b \hat{\Phi}^b-g^2\hat{\Phi}^a A_\mu^b A_\mu^b+
     2g\epsilon^{abc}A_\mu^b \partial_\mu \hat{\Phi}^c+g\epsilon^{abc}\partial_\mu A_\mu^b \hat{\Phi}^c\\+\lambda\Phi_0^2\hat{\Phi}^a)|\Phi|
    +2(\partial_\mu \hat{\Phi}^a
    +g\epsilon^{abc}A_\mu^b \hat{\Phi}^c)\partial_\mu |\Phi|+\hat{\Phi}^a\partial_\mu \partial_\mu |\Phi|
    \, ,
    \label{eq:expanded_higgs_eom}
\end{multline}
and the $SU(2)$ gauge field equations of motion are
\begin{multline}
    - |\Phi|^2 g \epsilon^{abc} \hat{\Phi}^b \partial_\mu \hat{\Phi}^c + |\Phi|^2 g^2 \hat{\Phi}^b A_\mu^b \hat{\Phi}^a - |\Phi|^2 g^2 A_\mu^a 
    =\partial_\nu \partial_\mu A_\nu^a - \partial_\nu \partial_\nu A_\mu^a \\
    + 2g \epsilon^{abc}A_\nu^c \partial_\nu A_\mu^b + g \epsilon^{abc} A_\mu^b \partial_\nu A_\nu^c\\
    + g\epsilon^{abc} A_\nu^b \partial_\mu A_\nu^c + g^2 A^b_\nu A^b_\nu A_\mu^a - g^2 A_\nu^b A_\mu^b A_\nu^a \,
\end{multline}
We solve these equations by discretizing the Euclidean Laplacian
\begin{multline}
\label{eq:discrete_laplacian}
    \partial_\mu\partial_\mu f_{ijkl} \simeq \frac{1}{\mathrm{d}x^2}\bigg[f_{ijk,l+1}+f_{ijk,l-1}+f_{ij,k+1,l}+f_{ij,k-1,l}
    \\+f_{i,j+1,kl}+f_{i,j-1,kl}+f_{i+1,jkl} + f_{i-1,jkl} - 8 f_{ijkl}\bigg] \, ,
\end{multline}
where $\dd x^2$ is the square step size of the grid (which we take to be uniform). We iteratively update the fields $f_{ijkl}$ via
\begin{multline} 
\label{eq:relaxation_step}
    f^\mathrm{new}_{ijkl} = f^\mathrm{old}_{ijkl} + \alpha\bigg[\frac{1}{8}\Big(f^\mathrm{old}_{ijk,l+1}+f^\mathrm{old}_{ijk,l-1}+f^\mathrm{old}_{ij,k+1,l}+f^\mathrm{old}_{ij,k-1,l}+f^\mathrm{old}_{i,j+1,kl}\\
    +f^\mathrm{old}_{i,j-1,kl}+f^\mathrm{old}_{i+1,jkl}
    + f^\mathrm{old}_{i-1,jkl}-\mathrm{d}x^2 g(f^\mathrm{old},\partial_\mu f^\mathrm{old}, \partial_\mu\partial_\nu f^\mathrm{old})\Big)-f^\mathrm{old}_{ijkl}\bigg] \, .
\end{multline}
where $\alpha$ is a step size parameter, and where $g$ is a functional representing all terms in the equations of motion beside the Laplacian. Eq.~\eqref{eq:relaxation_step} describes the step taken in each iteration -- it simply corresponds to plugging the fields into the right hand side of the equations of motion (with the $f_{ijkl}$ component of the discretized Laplacian, given by eq.~\eqref{eq:discrete_laplacian}, isolated on the left hand side.) and using this to compute a new field value $f_{ijkl}^\mathrm{new}$ at every lattice site.

Our initial guess for all three smooth field profiles $H,J,K$ takes the form
\begin{equation}
    \frac{H(u,v)}{\Phi_0}\Big|_\mathrm{init}=J(u,v)\Big|_\mathrm{init}=K(u,v)\Big|_\mathrm{init}=1-\exp\left[-\frac{(u\Phi_0)^2+([v-R]\Phi_0)^2}{2}\right] \, ,
\end{equation}
which correctly captures the limiting behavior on the loop and at infinity.
\section{The Higgsed BPST Instanton and the Hopf Map}\label{app:bpst_hopf_map}
In this appendix, we explicitly show that turning on an adjoint Higgs field on an $SU(2)$ BPST instanton background in regular gauge necessitates that the Higgs isospin defines a Hopf map at spacetime infinity. We start with the BPST instanton gauge field field configuration in regular gauge. We utilize matrix notation where the rows correspond to the isospin index $a$ and the columns correspond to the Euclidean spacetime index $\mu$. In this notation, the BPST instanton configuration at spatial infinity is defined by~\cite{Shifman:2012zz}
\begin{equation}
    \lim_{u,\, v\rightarrow\infty}A_\mu^a = \frac{2}{u^2+v^2}
    \begin{pmatrix}
    v\sin{\tau} & v\cos{\tau} & -u\sin{\varphi} & -u\cos{\varphi} \\
    -v \cos{\tau} & v \sin{\tau} & u\cos{\varphi} & -u\sin{\varphi} \\
    u \sin{\varphi} & -u\cos{\varphi} &  v\sin{\tau} & -v\cos{\tau}
    \end{pmatrix} \, .
\end{equation}
This configuration is pure gauge at infinity, meaning it can be written
\begin{equation}
    A_\mu^a T^a = i\Omega\partial_\mu \Omega^\dagger \, ,
\end{equation}
where $T^a=\sigma^a/2$ are the $SU(2)$ generators and where $\Omega$ is given by
\begin{equation}
    \Omega = \frac{1}{r}\left(x_4+i\sigma_i x_i\right) \, ,
\end{equation}
and $r=\sqrt{x_1^2+x_2^2+x_3^2+x_4^2}$. $\Omega$ is a textbook example of a $\mathrm{winding\;\,number}=1$ gauge transformation that takes $A_\mu^a$ from the topologically trivial sector to a sector that is non-trivially wound at infinity.\footnote{In the opposite convention where $\Omega$ and $\Omega^\dagger$ are flipped, the appropriate transformation is $\Omega=(x_4-i\sigma_ix_i)/r$ and a $-i$ appears in the pure gauge definition of $A_\mu$.} If we now fix this gauge field configuration and turn on a Higgs field $\Phi^a$, for the field configuration to have finite action, $A_\mu^a$ and $\Phi^a$ must be gauge matched at infinity so that $(D_\mu \Phi)^a =0$. This means that the Higgs must be configured as
\begin{align}
    \lim_{u,\, v\rightarrow\infty}\hat{\Phi}& = \Omega\frac{\sigma^3}{2}\Omega^\dagger \, , \\
    & = \frac{1}{r^2}
    \begin{pmatrix}
    2x_1x_3-2x_2x_4 \\
    2x_2x_3 + 2x_1x_4 \\
    -x_1^2-x_2^2+x_3^2+x_4^2
    \end{pmatrix} \, .
\end{align}
But this is a textbook example of a $S^3\rightarrow S^2$ Hopf map~\cite{Nakahara:2003nw}! Now, converting this map to the double azimuthal coordinates defined by eq.~\eqref{eq:coords_dp}, we obtain
\begin{align}
    \lim_{u,\, v\rightarrow\infty}\Phi^a &= \frac{1}{u^2+v^2}
    \begin{pmatrix}
    2uv\left(\cos\varphi\cos\tau - \sin\varphi\sin\tau\right) \\
    2uv(\sin\varphi\cos\tau + \cos\varphi\sin\tau) \\
    -u^2+v^2
    \end{pmatrix} \, , \\
    &=
    \begin{pmatrix}
    \sin\beta\cos(\varphi+\tau)\\
    \sin\beta\sin(\varphi+\tau)\\
    \cos\beta
    \end{pmatrix} \, .
\label{eq:bpst-hopf-map}
\end{align}
But this is exactly the Higgs isospin map defined by eq.~\eqref{eq:higgs_isospin} for $m=1$, $\ell=1$ in the $R\rightarrow 0$ limit. That is, our $SU(2)$ dyon loop Higgs configuration at infinity corresponds to that of a Higgs field on a BPST instanton background! We can now also obtain the ``Abelian'' part of $A_\mu^a$ at infinity via projecting onto the Higgs
\begin{equation}
    \lim_{u,\, v\rightarrow\infty}\hat{\Phi}^a A_\mu^a = \frac{2}{u^2+v^2}\begin{pmatrix}
        -u\sin{\varphi}\\
        u\cos{\varphi}\\
        v\sin{\tau}\\
        -v\cos{\tau}
    \end{pmatrix} \, ,
\label{eq:bpst-little-a}
\end{equation} 
where the ``Abelian'' field component is defined as $A^3$ in singular gauge. But notice that this is exactly the $\ell=1$ ``winding'' $U(1)$ map that appears in eq.~\eqref{eq:littlea}.  Hence, the twisting dyon topology is exactly equivalent to the topology of a Higgsed BPST instanton at spacetime infinity. Altogether, we can rewrite the asymptotic configuration of the instanton as
\begin{equation}
    \lim_{u^2+v^2\rightarrow\infty} A_\mu^a = -\epsilon^{abc}\hat{\Phi}^b\partial_\mu \hat{\Phi}^c + a_\mu \hat{\Phi}^a \, ,
\end{equation}
which follows from the fact that $D_\mu\hat{\Phi}^a=0$ at infinity. This gives us an appropriate instantonic boundary condition for our $\mathbb{R}^4$ dyon loop in \sec{sec:non-abelian-loop}.

\section{Uniqueness of BPST Instanton Solution in the Georgi-Glashow Model}\label{app:no-other-action-minimum}

In this appendix, we prove that the size-zero BPST instanton is the unique extremum of the $D=4$ Euclidean Georgi-Glashow action in the instanton number = $1$ topological sector. To do this, we rely on field rescaling arguments from Derrick's theorem~\cite{Derrick:1964ww,Weinberg:2012pjx} and the uniqueness of the BPST instanton as a solution to the (Higgsless) Euclidean Yang-Mills equations~\cite{Atiyah:1977pw}. First, we write the Euclidean Georgi-Glashow action in $D$ dimensions as
\begin{align}
    S &= 
    S_G + S_{D\Phi} + S_{V(\Phi)} \, , 
\end{align}
where
\begin{subequations}
\label{eq:action_terms_derrick}
\begin{align}
    S_G[A] &= \int \dd^D x \frac{1}{4g^2} G^{a}_{\mu \nu} G^{a}_{\mu \nu} \, ,\\[5pt]
    S_{D\Phi}[A,\Phi] & = \int \dd^D x\frac{1}{2} D_\mu \Phi^a D_\mu \Phi^a \, ,\\[5pt]
    S_{V(\Phi)}[\Phi] &=\int \dd^D x\frac{\lambda}{4} \left(\Phi^a \Phi^a - \Phi_0^2 \right)^2 \, .
\end{align}
\end{subequations}
Suppose we have found a field configuration $\{\bar{A}^a_\mu(x)\, ,\bar{\Phi}^a(x)\}$ that solves the Euclidean equations of motion. Now, consider the more general class of rescaled field configurations $\{\alpha\bar{A}^a_\mu(\alpha x),$ $\bar{\Phi}^a(\alpha x)\}$ where $\alpha\in\mathbb{R}^+$ is some rescaling parameter. The action associated with this more general class of field configurations can be related to that of the solution via
\begin{subequations}
\label{eq:action_terms_unscaled}
\begin{align}
    S_G(\alpha) &=\alpha^{(4-D)} S_G[\bar{A}] \, ,\\[5pt]
    S_{D\Phi}(\alpha) & = \alpha^{(2-D)}S_{D\Phi}[\bar{A},\bar{\Phi}] \, ,\\[5pt]
    S_{V(\Phi)}(\alpha) &=\alpha^{-D}S_{V(\Phi)}[\bar{\Phi}] \, .
\end{align}
\end{subequations}
Since a rescaling by $\alpha$ is a continuous deformation of the $\alpha=1$ field configuration, $\alpha=1$ is an extremum of the action if
\begin{equation}
    0=(D-4)S_G + (D-2) S_{D\Phi} + D S_{V(\Phi)} \, .
\end{equation}
But in Euclidean signature, $S_G,\, S_{D\Phi}, \, S_{V(\Phi)} \geq 0$, so in $D=4$, the above relation is only satisfied if $(D_\mu\Phi)^a=0$ and $V(\Phi)=0$ everywhere. This means that only the gauge kinetic term can contribute to the action non-trivially. The Bogomol'nyi bound~\cite{Bogomolny:1975de} constrains this term to be
\begin{equation}
\label{eq:bogomolnyi-bound}
    S_G \geq \frac{8\pi^2}{g^2} \, ,
\end{equation}
in the instanton number = 1 topological sector. In pure Yang-Mills, the standard BPST instanton is the unique solution the Euclidean equations of motion in this sector~\cite{Atiyah:1977pw}, and it exactly saturates the bound~\eqref{eq:bogomolnyi-bound}. It is usually endowed with an arbitrary scale parameter $\rho$ due to semiclassical scale invariance. However, in the Georgi-Glashow model, for $(D_\mu\Phi)^a$ to vanish everywhere, we have the additional constraint that we must take $\rho\rightarrow 0$, pushing the solution to the edge of configuration space where the entire Lagrangian density is concentrated at a point. Thus, the BPST instanton in the limit where its size is taken to zero is the unique solution to the Euclidean equations of motion in the Georgi-Glashow model, in agreement with our findings in~\sec{sec:relaxation} that there are no local minima associated with a smooth closed dyon loop, and that a minimum is only reached by taking the loop radius $R\rightarrow 0$. As we discuss in~\sec{sec:relaxation}, this limit exactly corresponds to the $\rho\rightarrow 0$ limit of the BPST instanton.

\bibliographystyle{JHEP}
\bibliography{draft}

\providecommand{\href}[2]{#2}\begingroup\raggedright\begin{thebibliography}{10}

\bibitem{Dirac:1931kp}
P.~A.~M. Dirac, \emph{{Quantised singularities in the electromagnetic field}}, \href{http://dx.doi.org/10.1098/rspa.1931.0130}{\emph{Proc. Roy. Soc. Lond. A} {\bf 133} (1931) 60--72}.

\bibitem{tHooft:1974kcl}
G.~'t~Hooft, \emph{{Magnetic Monopoles in Unified Gauge Theories}}, \href{http://dx.doi.org/10.1016/0550-3213(74)90486-6}{\emph{Nucl. Phys. B} {\bf 79} (1974) 276--284}.

\bibitem{Polyakov:1974ek}
A.~M. Polyakov, \emph{{Particle Spectrum in Quantum Field Theory}}, {\emph{JETP Lett.} {\bf 20} (1974) 194--195}.

\bibitem{Polchinski:2003bq}
J.~Polchinski, \emph{{Monopoles, duality, and string theory}}, \href{http://dx.doi.org/10.1142/S0217751X0401866X}{\emph{Int. J. Mod. Phys. A} {\bf 19S1} (2004) 145--156}, [\href{https://arxiv.org/abs/hep-th/0304042}{{\tt hep-th/0304042}}].

\bibitem{Kibble:1976sj}
T.~W.~B. Kibble, \emph{{Topology of Cosmic Domains and Strings}}, \href{http://dx.doi.org/10.1088/0305-4470/9/8/029}{\emph{J. Phys. A} {\bf 9} (1976) 1387--1398}.

\bibitem{Witten:1979ey}
E.~Witten, \emph{{Dyons of Charge e theta/2 pi}}, \href{http://dx.doi.org/10.1016/0370-2693(79)90838-4}{\emph{Phys. Lett. B} {\bf 86} (1979) 283--287}.

\bibitem{Agasian:2014uua}
N.~O. Agasian, \emph{{Instanton in the Georgi-Glashow model}}, \href{http://dx.doi.org/10.1134/S106377881408002X}{\emph{Phys. Atom. Nucl.} {\bf 77} (2014) 1181--1185}.

\bibitem{Fan:2021ntg}
J.~Fan, K.~Fraser, M.~Reece and J.~Stout, \emph{{Axion Mass from Magnetic Monopole Loops}}, \href{http://dx.doi.org/10.1103/PhysRevLett.127.131602}{\emph{Phys. Rev. Lett.} {\bf 127} (2021) 131602}, [\href{https://arxiv.org/abs/2105.09950}{{\tt 2105.09950}}].

\bibitem{Marciano:1976as}
W.~J. Marciano and H.~Pagels, \emph{{Chiral Charge Conservation and Gauge Fields}}, \href{http://dx.doi.org/10.1103/PhysRevD.14.531}{\emph{Phys. Rev. D} {\bf 14} (1976) 531}.

\bibitem{Christ:1979iw}
N.~H. Christ and R.~Jackiw, \emph{{Equality of Magnetic Charge and Pontryagin Index for {Yang-Mills} Dyons}}, \href{http://dx.doi.org/10.1016/0370-2693(80)90438-4}{\emph{Phys. Lett. B} {\bf 91} (1980) 228--232}.

\bibitem{Kim:1980rv}
Y.~Kim, I.-G. Koh, Y.-J. Park and S.-H. Yoon, \emph{{The Spatial Volume Integral of Tr F F for Dyons}}, {\emph{J. Korean Phys. Soc.} {\bf 13} (1980) 1--6}.

\bibitem{Rossi:1978qe}
P.~Rossi, \emph{{Propagation Functions in the Field of a Monopole}}, \href{http://dx.doi.org/10.1016/0550-3213(79)90163-9}{\emph{Nucl. Phys. B} {\bf 149} (1979) 170--188}.

\bibitem{Rossi:1982fq}
P.~Rossi, \emph{{EXACT RESULTS IN THE THEORY OF NONABELIAN MAGNETIC MONOPOLES}}, \href{http://dx.doi.org/10.1016/0370-1573(82)90081-3}{\emph{Phys. Rept.} {\bf 86} (1982) 317--362}.

\bibitem{Taubes:1984scl}
C.~H. Taubes, \emph{{Morse Theory and Monopoles: Topology in Long Range Forces}}, {\emph{NATO Sci. Ser. B} {\bf 115} (1984) 563--587}.

\bibitem{Jahn:1999wx}
O.~Jahn, \emph{{Instantons and monopoles in general Abelian gauges}}, \href{http://dx.doi.org/10.1088/0305-4470/33/15/307}{\emph{J. Phys. A} {\bf 33} (2000) 2997--3019}, [\href{https://arxiv.org/abs/hep-th/9909004}{{\tt hep-th/9909004}}].

\bibitem{Bruckmann:2000zs}
F.~Bruckmann, \emph{{Hopf defects as seeds for monopole loops}}, \href{http://dx.doi.org/10.1088/1126-6708/2001/08/030}{\emph{JHEP} {\bf 08} (2001) 030}, [\href{https://arxiv.org/abs/hep-th/0011249}{{\tt hep-th/0011249}}].

\bibitem{Bruckmann:2002jm}
F.~Bruckmann, \emph{{Monopoles from instantons}},  in \emph{{NATO Advanced Research Workshop on Confinement, Topology, and other Nonperturbative Aspects of QCD}}, 4, 2002.
\newblock \href{https://arxiv.org/abs/hep-th/0204241}{{\tt hep-th/0204241}}.

\bibitem{Saurabh:2017ryg}
A.~Saurabh and T.~Vachaspati, \emph{{Monopole-antimonopole Interaction Potential}}, \href{http://dx.doi.org/10.1103/PhysRevD.96.103536}{\emph{Phys. Rev. D} {\bf 96} (2017) 103536}, [\href{https://arxiv.org/abs/1705.03091}{{\tt 1705.03091}}].

\bibitem{Chernodub:1995tt}
M.~N. Chernodub and F.~V. Gubarev, \emph{{Instantons and monopoles in maximal Abelian projection of SU(2) gluodynamics}}, {\emph{JETP Lett.} {\bf 62} (1995) 100--104}, [\href{https://arxiv.org/abs/hep-th/9506026}{{\tt hep-th/9506026}}].

\bibitem{Hart:1995wk}
A.~Hart and M.~Teper, \emph{{Instantons and monopoles in the maximally Abelian gauge}}, \href{http://dx.doi.org/10.1016/0370-2693(96)00017-2}{\emph{Phys. Lett. B} {\bf 371} (1996) 261--269}, [\href{https://arxiv.org/abs/hep-lat/9511016}{{\tt hep-lat/9511016}}].

\bibitem{Brower_1997}
R.~C. Brower, K.~N. Orginos and C.-I. Tan, \emph{Magnetic monopole loop for the yang-mills instanton}, \href{http://dx.doi.org/10.1103/physrevd.55.6313}{\emph{Physical Review D} {\bf 55} (May, 1997) 6313–6326}.

\bibitem{Bornyakov:1997at}
V.~Bornyakov and G.~Schierholz, \emph{{Instantons are dyon loops}}, \href{http://dx.doi.org/10.1016/S0920-5632(96)00694-9}{\emph{Nucl. Phys. B Proc. Suppl.} {\bf 53} (1997) 484--487}.

\bibitem{Csaki:2024ajo}
C.~Cs\'aki, R.~Ovadia, O.~Telem, J.~Terning and S.~Yankielowicz, \emph{{Abelian instantons and monopole scattering}}, \href{http://dx.doi.org/10.1007/JHEP11(2024)165}{\emph{JHEP} {\bf 11} (2024) 165}, [\href{https://arxiv.org/abs/2406.13738}{{\tt 2406.13738}}].

\bibitem{Chen:2025buv}
S.~Chen, A.~Cherman and M.~Neuzil, \emph{{Symmetry theta angles and topological Witten effects}},  \href{https://arxiv.org/abs/2507.00220}{{\tt 2507.00220}}.

\bibitem{Adler:1969gk}
S.~L. Adler, \emph{{Axial vector vertex in spinor electrodynamics}}, \href{http://dx.doi.org/10.1103/PhysRev.177.2426}{\emph{Phys. Rev.} {\bf 177} (1969) 2426--2438}.

\bibitem{Bell:1969ts}
J.~S. Bell and R.~Jackiw, \emph{{A PCAC puzzle: $\pi^0 \to \gamma \gamma$ in the $\sigma$ model}}, \href{http://dx.doi.org/10.1007/BF02823296}{\emph{Nuovo Cim. A} {\bf 60} (1969) 47--61}.

\bibitem{Choi:2022fgx}
Y.~Choi, H.~T. Lam and S.-H. Shao, \emph{{Non-invertible Gauss law and axions}}, \href{http://dx.doi.org/10.1007/JHEP09(2023)067}{\emph{JHEP} {\bf 09} (2023) 067}, [\href{https://arxiv.org/abs/2212.04499}{{\tt 2212.04499}}].

\bibitem{PhysRev.177.2426}
S.~L. Adler, \emph{Axial-vector vertex in spinor electrodynamics}, \href{http://dx.doi.org/10.1103/PhysRev.177.2426}{\emph{Phys. Rev.} {\bf 177} (Jan, 1969) 2426--2438}.

\bibitem{Bell:348417}
J.~S. Bell and R.~W. Jackiw, \emph{{A PCAC puzzle: $\pi ^{0} \rightarrow \gamma \gamma $ in the $\sigma $-model}}, \href{http://dx.doi.org/10.1007/BF02823296}{\emph{Nuovo Cimento} {\bf 60} (1969) 47--61}.

\bibitem{shao2024whatsundonetasilectures}
S.-H. Shao, \emph{What's done cannot be undone: Tasi lectures on non-invertible symmetries},  2024.

\bibitem{Cordova:2022ieu}
C.~Cordova and K.~Ohmori, \emph{{Noninvertible Chiral Symmetry and Exponential Hierarchies}}, \href{http://dx.doi.org/10.1103/PhysRevX.13.011034}{\emph{Phys. Rev. X} {\bf 13} (2023) 011034}, [\href{https://arxiv.org/abs/2205.06243}{{\tt 2205.06243}}].

\bibitem{Affleck:1980mp}
I.~Affleck, \emph{{On Constrained Instantons}}, \href{http://dx.doi.org/10.1016/0550-3213(81)90307-2}{\emph{Nucl. Phys. B} {\bf 191} (1981) 429}.

\bibitem{Reece:2023czb}
M.~Reece, \emph{{TASI Lectures: (No) Global Symmetries to Axion Physics}}, \href{http://dx.doi.org/10.22323/1.439.0008}{\emph{PoS} {\bf TASI2022} (2024) 008}, [\href{https://arxiv.org/abs/2304.08512}{{\tt 2304.08512}}].

\bibitem{Wu:1975es}
T.~T. Wu and C.~N. Yang, \emph{{Concept of Nonintegrable Phase Factors and Global Formulation of Gauge Fields}}, \href{http://dx.doi.org/10.1103/PhysRevD.12.3845}{\emph{Phys. Rev. D} {\bf 12} (1975) 3845--3857}.

\bibitem{Tong:2005un}
D.~Tong, \emph{{TASI lectures on solitons: Instantons, monopoles, vortices and kinks}},  in \emph{{Theoretical Advanced Study Institute in Elementary Particle Physics}: {Many Dimensions of String Theory}}, 6, 2005.
\newblock \href{https://arxiv.org/abs/hep-th/0509216}{{\tt hep-th/0509216}}.

\bibitem{Shifman:2012zz}
M.~Shifman, \emph{{Advanced topics in quantum field theory.}: {A lecture course}}.
\newblock Cambridge Univ. Press, Cambridge, UK, 2, 2012, \href{http://dx.doi.org/10.1017/9781108885911}{10.1017/9781108885911}.

\bibitem{Weinberg:2012pjx}
E.~J. Weinberg, \emph{{Classical solutions in quantum field theory}: {Solitons and Instantons in High Energy Physics}}.
\newblock Cambridge Monographs on Mathematical Physics. Cambridge University Press, 9, 2012, \href{http://dx.doi.org/10.1017/CBO9781139017787}{10.1017/CBO9781139017787}.

\bibitem{fox1945torus}
R.~H. Fox, \emph{Torus homotopy groups}, {\emph{Proceedings of the National Academy of Sciences} {\bf 31} (1945) 71--74}.

\bibitem{fox1948homotopy}
R.~H. Fox, \emph{Homotopy groups and torus homotopy groups}, {\emph{Annals of Mathematics} {\bf 49} (1948) 471--510}.

\bibitem{tHooft:1976snw}
G.~'t~Hooft, \emph{{Computation of the Quantum Effects Due to a Four-Dimensional Pseudoparticle}}, \href{http://dx.doi.org/10.1103/PhysRevD.14.3432}{\emph{Phys. Rev. D} {\bf 14} (1976) 3432--3450}.

\bibitem{Affleck:1981ag}
I.~K. Affleck and N.~S. Manton, \emph{{Monopole Pair Production in a Magnetic Field}}, \href{http://dx.doi.org/10.1016/0550-3213(82)90511-9}{\emph{Nucl. Phys. B} {\bf 194} (1982) 38--64}.

\bibitem{Julia:1975ff}
B.~Julia and A.~Zee, \emph{{Poles with Both Magnetic and Electric Charges in Nonabelian Gauge Theory}}, \href{http://dx.doi.org/10.1103/PhysRevD.11.2227}{\emph{Phys. Rev. D} {\bf 11} (1975) 2227--2232}.

\bibitem{Shnir:2005vvi}
Y.~M. Shnir, \emph{{Magnetic Monopoles}}.
\newblock Text and Monographs in Physics. Springer, Berlin/Heidelberg, 2005, \href{http://dx.doi.org/10.1007/3-540-29082-6}{10.1007/3-540-29082-6}.

\bibitem{Garcia-Valdecasas:2024cqn}
E.~Garc\'\i{}a-Valdecasas, M.~Reece and M.~Suzuki, \emph{{Monopole Breaking of Chern-Weil Symmetries}},  \href{https://arxiv.org/abs/2408.00067}{{\tt 2408.00067}}.

\bibitem{Arafune:1974uy}
J.~Arafune, P.~G.~O. Freund and C.~J. Goebel, \emph{{Topology of Higgs Fields}}, \href{http://dx.doi.org/10.1063/1.522518}{\emph{J. Math. Phys.} {\bf 16} (1975) 433}.

\bibitem{Bais:1976fr}
F.~A. Bais, \emph{{SO(3) Monopoles and Dyons with Multiple Magnetic Charge}}, \href{http://dx.doi.org/10.1016/0370-2693(76)90123-4}{\emph{Phys. Lett. B} {\bf 64} (1976) 465--468}.

\bibitem{Nakahara:2003nw}
M.~Nakahara, \emph{{Geometry, topology and physics}}.
\newblock 2003.

\bibitem{Cho:1980nx}
Y.~M. Cho, \emph{{Extended Gauge Theory and Its Mass Spectrum}}, \href{http://dx.doi.org/10.1103/PhysRevD.23.2415}{\emph{Phys. Rev. D} {\bf 23} (1981) 2415}.

\bibitem{Faddeev:1998eq}
L.~D. Faddeev and A.~J. Niemi, \emph{{Partially dual variables in SU(2) Yang-Mills theory}}, \href{http://dx.doi.org/10.1103/PhysRevLett.82.1624}{\emph{Phys. Rev. Lett.} {\bf 82} (1999) 1624--1627}, [\href{https://arxiv.org/abs/hep-th/9807069}{{\tt hep-th/9807069}}].

\bibitem{EMGR_upcoming}
I.~Garcia~Garcia and E.~Maderazo, \emph{{Abelian Instantons in Einstein-Maxwell Theory (in preparation)}},  2025.

\bibitem{Fukuda:2020imw}
H.~Fukuda and K.~Yonekura, \emph{{Witten effect, anomaly inflow, and charge teleportation}}, \href{http://dx.doi.org/10.1007/JHEP01(2021)119}{\emph{JHEP} {\bf 01} (2021) 119}, [\href{https://arxiv.org/abs/2010.02221}{{\tt 2010.02221}}].

\bibitem{Heidenreich:2020pkc}
B.~Heidenreich, J.~McNamara, M.~Montero, M.~Reece, T.~Rudelius and I.~Valenzuela, \emph{{Chern-Weil global symmetries and how quantum gravity avoids them}}, \href{http://dx.doi.org/10.1007/JHEP11(2021)053}{\emph{JHEP} {\bf 11} (2021) 053}, [\href{https://arxiv.org/abs/2012.00009}{{\tt 2012.00009}}].

\bibitem{harris2020array}
C.~R. Harris, K.~J. Millman, S.~J. van~der Walt, R.~Gommers, P.~Virtanen, D.~Cournapeau et~al., \emph{Array programming with {NumPy}}, \href{http://dx.doi.org/10.1038/s41586-020-2649-2}{\emph{Nature} {\bf 585} (Sept., 2020) 357--362}.

\bibitem{2020SciPy-NMeth}
P.~Virtanen, R.~Gommers, T.~E. Oliphant, M.~Haberland, T.~Reddy, D.~Cournapeau et~al., \emph{{{SciPy} 1.0: Fundamental Algorithms for Scientific Computing in Python}}, \href{http://dx.doi.org/10.1038/s41592-019-0686-2}{\emph{Nature Methods} {\bf 17} (2020) 261--272}.

\bibitem{jax2018github}
J.~Bradbury, R.~Frostig, P.~Hawkins, M.~J. Johnson, C.~Leary, D.~Maclaurin et~al., \emph{{JAX}: composable transformations of {P}ython+{N}um{P}y programs},  2018.

\bibitem{Derrick:1964ww}
G.~H. Derrick, \emph{{Comments on nonlinear wave equations as models for elementary particles}}, \href{http://dx.doi.org/10.1063/1.1704233}{\emph{J. Math. Phys.} {\bf 5} (1964) 1252--1254}.

\bibitem{Atiyah:1977pw}
M.~F. Atiyah and R.~S. Ward, \emph{{Instantons and Algebraic Geometry}}, \href{http://dx.doi.org/10.1007/BF01626514}{\emph{Commun. Math. Phys.} {\bf 55} (1977) 117--124}.

\bibitem{Bogomolny:1975de}
E.~B. Bogomolny, \emph{{Stability of Classical Solutions}}, {\emph{Sov. J. Nucl. Phys.} {\bf 24} (1976) 449}.

\end{thebibliography}\endgroup

\end{document}